%% file: Geminga_final.tex
\documentclass[longauth]{aa-package/aa}
\usepackage{url,hyperref}
\usepackage{footmisc}
\usepackage{epstopdf}
\usepackage[varg]{txfonts}
\usepackage[percent]{overpic} 
\usepackage{booktabs}
\usepackage{relsize}
\usepackage{natbib}
\usepackage{xcolor}
\usepackage[switch]{lineno}

\makeatletter
\renewcommand*{\@fnsymbol}[1]{\ifcase#1\or*\or$\dagger$\or$\ddagger$\or**\or$\dagger\dagger$\or$\ddagger\ddagger$\fi}
\makeatother

\usepackage{amstext}

\begin{document}


\title{Detection of extended $\gamma$-ray emission around the Geminga pulsar with H.E.S.S.}

\include{./authors_Detection_of_GemingaHESS_AA}

\date{Received: 23/12/2022 -- Accepted: 04/04/2023}

\abstract{Geminga is an enigmatic radio-quiet $\gamma$-ray pulsar located at a mere 250\,pc distance from Earth. Extended very-high-energy $\gamma$-ray emission around the pulsar was discovered by Milagro and later confirmed by HAWC, which are both water Cherenkov detector-based experiments. However, evidence for the Geminga pulsar wind nebula in gamma rays has long evaded detection by imaging atmospheric Cherenkov telescopes (IACTs) despite targeted observations. The detection of $\gamma$-ray emission on angular scales $\gtrsim2^\circ$ poses a considerable challenge for the background estimation in IACT data analysis. With recent developments in understanding the complementary background estimation techniques of water Cherenkov and atmospheric Cherenkov instruments, the H.E.S.S. IACT array can now confirm the detection of highly extended $\gamma$-ray emission around the Geminga pulsar with a radius of at least $3^\circ$ in the energy range $0.5-40$\,TeV. 
We find no indications for statistically significant asymmetries or energy-dependent morphology. 
A flux normalisation of $(2.8\pm0.7)\times10^{-12}$\,cm$^{-2}$s$^{-1}$TeV$^{-1}$ at 1\,TeV is obtained within a $1^\circ$ radius region around the pulsar. To investigate the particle transport within the halo of energetic leptons around the pulsar, we fitted an electron diffusion model to the data. The normalisation of the diffusion coefficient obtained of $D_0 = 7.6^{+1.5}_{-1.2} \times 10^{27}$ cm$^2$s$^{-1}$, at an electron energy of 100\,TeV, is compatible with values previously reported for the pulsar halo around Geminga, which is considerably below the Galactic average. 
}


\maketitle

\makeatletter
\renewcommand*{\@fnsymbol}[1]{\ifcase#1\@arabic{#1}\fi}
\makeatother

\section{Introduction}
\label{sec:intro}

The Geminga pulsar (PSR\,J0633+1746) is a radio-quiet $\gamma$-ray source that was established as a pulsar in 1992, and at 250\,pc is one of the closest pulsars to Earth \citep{Bignami1983ApJ...272L...9B,Bertsch1992Natur.357..306B,GemingaPulsar1992Natur.357..287B,Faherty2007Ap&SS.308..225F}. Searches for extended $\gamma$-ray emission around Geminga have been conducted ever since, proving unsuccessful for many years \citep{1993A&A...274L..17A_whippleGeminga,1999A&A...346..913A_HegraGeminga}, until it was first detected at TeV energies by Milagro in 2007, with a diameter of $2.8^\circ \pm0.8^\circ$ \citep{Milagro07}. This discovery was subsequently confirmed by the High Altitude Water Cherenkov (HAWC) collaboration, presenting evidence for significantly extended emission on angular scales of up to $\sim5.5^\circ$ radius \citep{2HWC17}. However, a detection with imaging atmospheric Cherenkov telescopes (IACTs) remained elusive \citep{2009APh....32..120S_PACTgeminga,veritas2009F,Magic16}. 
With a spin-down luminosity of $\dot{E} = 3.2\times 10^{34}\,\mathrm{ergs}^{-1}$, a spin period of $P=237$\,ms and a characteristic age of $\tau_c = 342$\,kyr, Geminga is one of the oldest pulsars around which extended very-high-energy $\gamma$-ray emission has been detected, providing evidence of energetic electron acceleration by middle-aged pulsars ($\tau_c \sim 0.1 - 1$\,Myr) \citep{Manchester05}.\footnote{The term `electron' is used to refer, collectively, to electrons and positrons throughout, unless explicitly stated otherwise.} 

The morphology of the emission as seen with HAWC was found to indicate considerably slower diffusion than values typical for the interstellar medium (ISM), yielding diffusion coefficients a factor $\sim100$ lower at 100\,TeV \citep{HAWC17}. 
Geminga and the similar nearby pulsar PSR\,B0656+14 are considerably older pulsars, with longer periods and lower spin-down powers with respect to other pulsars associated with extended TeV $\gamma$-ray emission. 
It has been proposed that the $\gamma$-ray emission regions around these systems are in a different evolutionary stage to other TeV pulsar wind nebulae (PWNe) and they constitute a distinct source class of `TeV halos' or `pulsar halos' \citep{Giacinti2020,Linden2017PhRvD..96j3016L}.\footnote{We adopt the latter term `pulsar halo' to distinguish from the similar escape phenomenon occurring in other sources, such as halos around supernova remnants \citep{2021A&A...654A.139Brose}. }
Within such halos, the $\gamma$-ray emission is due to inverse Compton (IC) scattering by electrons that have escaped from the PWN and the energy density of these electrons is lower than the ISM energy density, that is they do not dominate the surrounding region dynamically or energetically. 
As such, in contrast to the many PWNe detected with the High Energy Stereoscopic System (H.E.S.S.), with 12 firmly identified in the H.E.S.S. Galactic Plane Survey \citep{HGPS}, the detection of extended $\gamma$-ray emission around the Geminga pulsar with H.E.S.S. constitutes the first unambiguous detection of a pulsar halo at TeV energies by IACTs. %

Extensive air showers (EAS) caused by cosmic rays are the primary background source of triggered events for a $\gamma$-ray analysis with IACTs. Despite state-of-the-art techniques in separating gamma-initiated from hadron-initiated EAS, there remains an irreducible background of gamma-like hadronic (mostly proton) showers -- those in which a large fraction of the energy is transferred to neutral pions in the first interaction  \citep{2007APh....28...72MaierKnapp}. 
As the rate of background events can vary based on the atmospheric conditions, observing direction, and hardware settings, methods to estimate and model the background level typically rely on counting gamma-like events within a region of the sky in the same dataset away from the target region (`Off'), and subtracting this level from the region of interest (`On') \citep{BergeFunkHinton07}.
In the case of the Geminga halo, the emission fills the field of view, such that there is no region free from $\gamma$-ray emission available for such a background estimation. This results in emission belonging to the halo being counted as background.

Previous IACT observations of the region have resulted in upper limits when probing angular scales in the range $0.1^\circ$ to $0.3^\circ$ \citep{veritas2009F,Magic16}. 
Within the X-ray range a PWN has been identified with two lateral tails of length $\sim2-3'$ and an axial tail of length $45''$, motivating searches on these angular scales \citep{Caraveo1345,Pavlov10,Posselt17}. 
Analysis and detection of emission on larger scales is challenging for IACTs, for reasons outlined above \citep{BergeFunkHinton07}. 
Dedicated analysis approaches developed for this sky region include matching observation conditions between observations of this region and of empty sky fields, in order to estimate the background across the full region of interest \citep{Flinders15,veritas2019ICRC...36..616A,Hona:2021oW}. 

As shown in a systematic study of differences between the analysis procedures of the water Cherenkov detector facility HAWC and the H.E.S.S. IACT array \citep{HAWCHESS_2021ApJ...917....6A}, the two experimental techniques are complementary approaches, with HAWC providing good sensitivity to high energies ($E \gtrsim 10$\,TeV) and extended emission regions. 
In contrast, H.E.S.S. can provide detailed morphological and spectral studies, thanks to its few-arcmin angular and 10-15\% energy resolution. 
Given the extent of the significant emission detected with HAWC around the Geminga pulsar (out to $\sim5^\circ$ radius), it is not currently possible for IACTs to measure the full extent of the emission using standard background estimation techniques. 
However, as we also show in the analysis presented here, it is nevertheless possible for IACTs to significantly detect degree-scale extended emission whilst controlling the estimated background, and to start probing the morphological and spectral properties of the inner regions of such sources \citep{2022A&A...666A.124A_West1}. 

Searches for spectral variation, energy-dependent morphology or asymmetries to the emission are of particular interest to investigate properties of the large scale pulsar halo. Similarly, investigating the transport of energetic particles in the vicinity of the pulsar is important to establish whether the properties are consistent with suppressed diffusion as seen in previous analyses.

\begin{figure}
    \centering
    \includegraphics[width=\columnwidth]{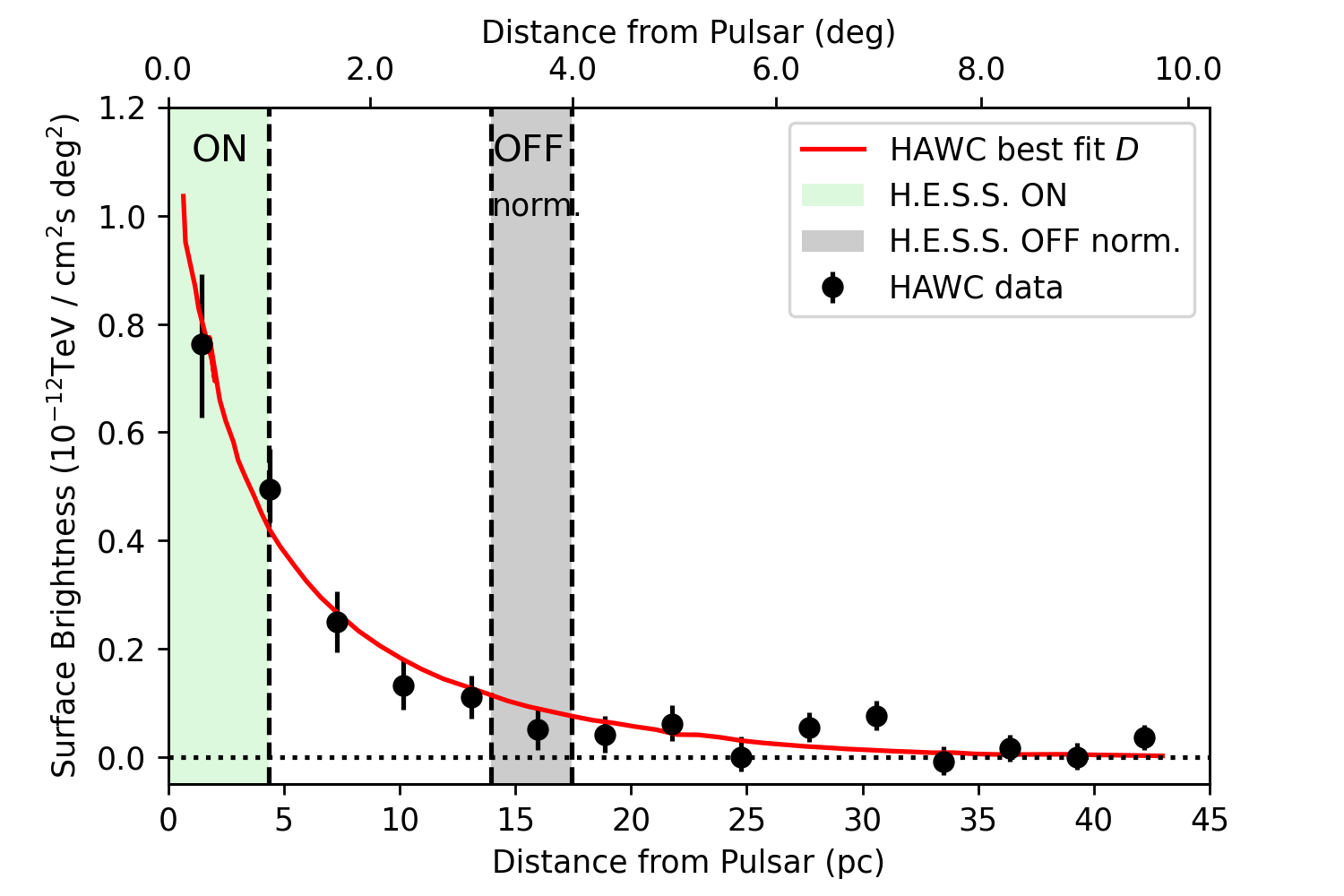}
    \caption{Surface brightness profile of emission around the Geminga pulsar measured with HAWC \citep{HAWC17}. Shaded regions indicate the parts of the emission profile that are used as ON (radius $\theta \lesssim 1.0^\circ$) and OFF normalisation (radius $\theta \gtrsim 3.2^\circ$) regions to estimate the background level and evaluate the significance in the analysis of the 2019 H.E.S.S. dataset. The region shown for normalisation of the OFF counts is only accessible with the 2019 dataset, due to the wider pointing strategy used. }
    \label{fig:hawc_onoff}
\end{figure}

\section{H.E.S.S. data and analysis}
\label{sec:hessdata}
\subsection{H.E.S.S. observations}
H.E.S.S. is an array of five IACTs, located in the Khomas Highland of Namibia at 1800\,m above sea level. Four of the telescopes (CT1-4) have mirror areas of 107\,m$^2$ whilst the fifth (CT5) has a mirror area of 612\,m$^2$ and correspondingly lower energy threshold \citep{Holler15ICRC}. Due to its smaller field of view, of $3.2^\circ$ (compared with the $5^\circ$ field of view for the smaller telescopes), CT5 is not used in this analysis. 
The electronics of the cameras of the CT1-4 telescopes were upgraded in 2017 \citep{Ashton2020}.
Observation data are collected with H.E.S.S. in `runs' typically 28\,minutes in length, during which telescopes are pointed towards a specific sky position. 

H.E.S.S. observed the Geminga region using `wobble' mode offsets of $0.5^\circ$ and $0.7^\circ$ around the Geminga pulsar (R.A. $06$h$33$m$54$s, Dec $+17^\circ 46'13''$) over two seasons in 2006 - 2008, for a total of 14.2 hours of livetime (see Table \ref{tab:data}) \citep{Aharonian06}. 
With these default wobble offsets, no $\gamma$-ray emission has been detected when using background estimation techniques probing radii of $\lesssim 0.3^\circ$. 
\begin{table}
\caption{Observation data taken on the Geminga region.}
\label{tab:data}
\begin{tabular}{ccccc}
Telescopes & Time period & Livetime & $\theta_z$ & Offset \\
\hline
CT1-4 & Nov 2006 & 7.7\,h & 42.2$^\circ$ & $\pm0.5^\circ$ \\
CT1-4 & Jan 2008 & 6.5\,h & 42.0$^\circ$ & $\pm0.7^\circ$ \\
CT1-4 & Jan-Mar 2019 & 27.2\,h & 43.5$^\circ$ & $\pm1.6^\circ$ \\
\end{tabular}
\tablefoot{
For each dataset we provide the cumulative livetime after run selection, average zenith angle $\theta_z$ and wobble offset in both R.A. and Dec. of the observation positions from the location of the Geminga pulsar.
}
\end{table}
Further observations were taken during the first quarter of 2019, employing a pointing strategy with much wider pointing offsets of $\pm1.6^\circ$ from the pulsar in R.A. and Dec., for a total livetime of 27.2 hours. These offsets are among the widest `wobble' mode pointing offsets yet used with IACTs, thereby pushing the limits of standard analysis techniques. 

\subsection{Analysis}

Considerable advances in gamma-hadron separation were achieved by the use of template or model-based approaches. In this analysis, a sensitive likelihood-based template fitting analysis is used \citep{Parsons14} and the results are cross-checked using an independent calibration and analysis chain \citep{deNaurois09}.

The Geminga region is observable with H.E.S.S. from November to April only at large zenith angles $> 40^\circ$, which raises the intrinsic energy threshold of the analysis. 
Additionally, depending on the method of background estimation and due to systematic uncertainties at the level of the statistical uncertainties or higher (caused for example by variation in the atmospheric conditions \citealt{HAHN201425}) we use a conservative set of selection cuts on reconstruction quality, known as \emph{hard} selection cuts, such that only the best candidate $\gamma$-ray events are retained. This also has the effect of improving the signal-to-background ratio such that the uncertainty on the background normalisation is reduced. 
Correspondingly, the energy threshold is 1\,TeV for the spectral analysis; whilst for the morphological analysis, we used an energy threshold of 0.5\,TeV. This is due to the higher event reconstruction quality and an energy bias of less than 10\% required for the spectral analysis.

\begin{figure*}
\centering
\includegraphics[width=\columnwidth]{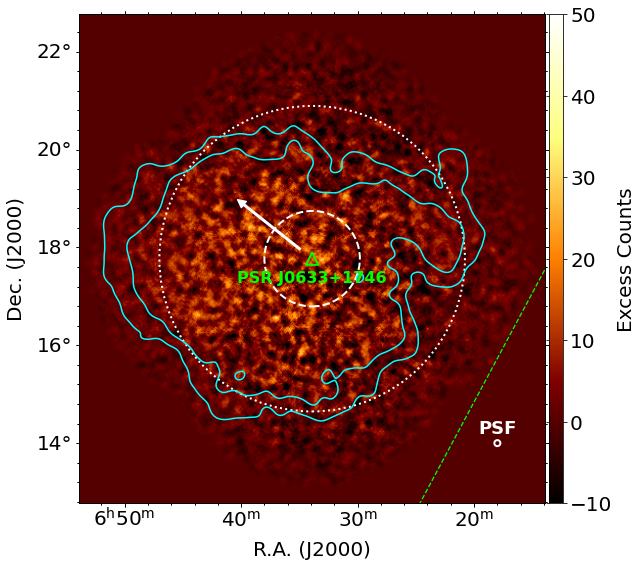}
\includegraphics[width=\columnwidth]{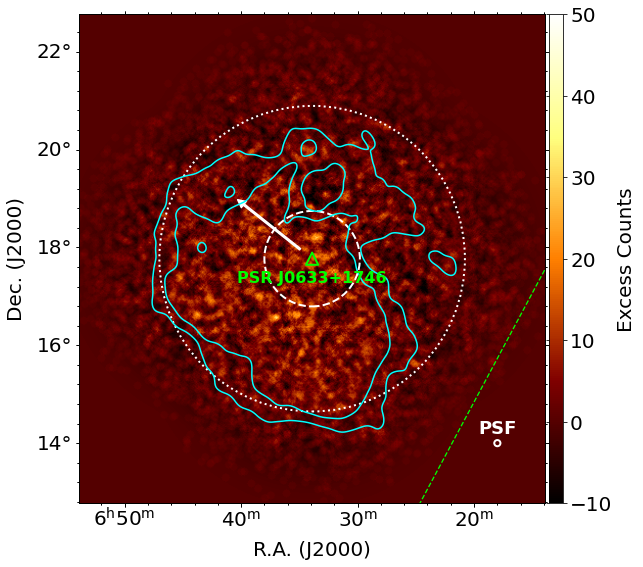}
\caption{Excess counts sky maps of the Geminga region above 0.5\,TeV using the 2019 dataset with two different background estimation methods. 
Left: On-Off background estimation method. Right: field-of-view background estimation method. The maps are over-sampled with a 0.08º correlation radius, with contours at the level of 50 and 100 excess counts from maps over-sampled with a 0.5º correlation radius. 
The 68\% containment PSF is shown for comparison and has a value of 0.06º, valid for the innermost $\theta <1^\circ$ region around the pulsar in which the significance is evaluated (indicated by a dashed circle). The Galactic plane is indicated by a green dashed line and the pulsar location with a green triangle, whilst an arrow indicates the proper motion direction of the pulsar (arbitrary length). Background normalisation is performed using regions at $\theta >3.2^\circ$, indicated by a dotted circle. }
\label{fig:maps}
\end{figure*}

\section{Detection of extended $\gamma$-ray emission}
\label{sec:detection}

A systematic study of analysis differences and background estimation techniques between H.E.S.S. and HAWC was conducted and applied to the Galactic plane \citep{HAWCHESS_2021ApJ...917....6A}. 
By adapting the analysis of H.E.S.S. data to techniques suitable for angular scales comparable to those probed by HAWC, it was found that the detectability of large, extended sources could be improved. This enabled the measurement of $\gamma$-ray emission from a very extended region and, although the background analysis remains challenging, the H.E.S.S. data retains the advantage of being able to identify smaller scale structures, including potential point-like sources.

As the $\gamma$-ray emission is considerably more extended than most other single TeV $\gamma$-ray sources detected by IACTs to date, a variety of approaches are used to verify the detection and nature of the emission. Particular care needed to be taken with the background estimation, which is described in detail in sections \ref{sec:background68} and \ref{sec:background19}. It is not yet possible to evaluate the true extent of the TeV $\gamma$-ray emission, as it extends beyond the sky region available with the current dataset. 

Using an integration region of $1^\circ$ radius, extended emission around the Geminga pulsar is detected at $> 6\sigma$ (evaluated using \citealt{1983ApJ...272..317LiMa}) in the 2006-2008 dataset and similarly at $\sim 9\sigma$ in the 2019 dataset.
Due to the differences in wobble offset between the 2006-2008 and the 2019 observations, different background estimation techniques are used for the two datasets. 

\begin{table}[]
    \caption{Li \& Ma significance for $\gamma$-ray emission within a $1^\circ$ radius around the Geminga pulsar obtained with different background methods and datasets. }
    \label{tab:signif}
    \centering
    \begin{tabular}{cccc}
    Dataset & Background method & $\sigma$ in $1^\circ$ & Cut level\\
    \hline
    2006-08 & Ring & $9\sigma$ & \emph{std} \\
    2006-08 & On-Off & $6.6\sigma$ & \emph{hard} \\    
    2019 & Field-of-View & $9.8\sigma$ & \emph{hard} \\
    2019 & On-Off (1) & $9.6\sigma$ & \emph{hard} \\
    2019 & On-Off (2) & $11.6\sigma$ & \emph{hard}
    \end{tabular}
    \tablefoot{
    For the Field-of-View background method, the Cash statistic was used to evaluate the significance instead. The level of selection cuts applied (\emph{std} or \emph{hard}) is also indicated.
    }
\end{table}

\subsection{Background estimation: 2006-2008 dataset}
\label{sec:background68}

Revisiting the Geminga observations taken with H.E.S.S. in 2006-2008, the detection of emission over a much larger angular scale is enabled by increasing the exclusion region\footnote{the region excluded from background estimation} radius from $\sim0.4^\circ$ to $1.5^\circ$. 
The reflected region background method is well suited to spectral analysis, in which a region of the same size as the ON region yet reflected across the telescope pointing direction is used to estimate the background. However, this is not possible for cases where the ON region size (here of $1^\circ$ radius) exceeds the pointing offset (here of $\pm0.5^\circ-0.7^\circ$, see Table \ref{tab:data}). Additionally, due to the large exclusion regions, a spectral analysis of the emission using the reflected background method is not possible on the Geminga region for any of the current H.E.S.S. datasets \citep{2001A&A...370..112A_hegraCasAreflbg}. 

Employing the ring background method with a fixed ring thickness of $0.5^\circ$ and a minimum radius in excess of $1.5^\circ$, for a 250\,pc distance to Geminga, this corresponds to background estimation from radii $>$6.5\,pc away from the pulsar \citep{BergeFunkHinton07}. As the background counts are sampled from the same region of the sky in the ring background method, standard (\emph{std}) selection cuts are used, whereas more conservative (\emph{hard}) selection cuts are used for other methods of background estimation. 
With such a ring background subtraction, H.E.S.S. is sensitive to the difference in $\gamma$-ray emission between the ON region at small radii from the pulsar, and the emission at larger radii that is used to estimate the background. 
The limited field of view prevents using a background ring radius large enough such that the background region does not contain emission from the source. The measured flux is therefore a relative measurement and will underestimate the true flux (see also \citealt{tevpageminga}).

Two different background methods, Ring and On-Off, are applied to evaluate the significance of $\gamma$-ray emission within a $1^\circ$ radius region around the pulsar in this dataset, with a consistent detection obtained using both approaches (see Table \ref{tab:signif}). 
When using the On-Off background estimation, the entire field of view of the observations is considered as the On region, with Off data taken from observing runs matched for comparable conditions. The parameters used to match the Off data to On data include the zenith angle of the observations, the run duration, and the combination of telescopes participating. The presence of all four telescopes CT1-4 is required for this analysis in order to provide a smooth acceptance\footnote{the probability of accepting a $\gamma$-ray candidate event reconstructed at a certain position and energy} across the sky region. 
The Off runs used to estimate the background are extragalactic observations from 2004-2009 with no significant source detected in the field of view. 

\subsection{Background estimation: 2019 dataset}
\label{sec:background19}

Following this detection, observations of the Geminga region were conducted in 2019 at large wobble offsets of $1.6^\circ$ around the pulsar, out to which the camera acceptance remains at $\gtrsim60\%$ of the on-axis acceptance \citep{Aharonian06}. 
 Once again, the emission from the region around Geminga is found to fill the available field of view, proving a considerable challenge for background estimation in the analysis. Therefore, two different methods are used; the so-called On-Off and field-of-view background approaches \citep{1989ApJ...342..379WeekesOnOff,BergeFunkHinton07}. 

In order to provide an estimate of the systematic uncertainties of the On-Off background estimator and cross-check our results, we used two independent matched lists of Off runs, with different tolerance levels in the matching criteria (see Table \ref{tab:matchoff}). Both lists are comprised of observations from 2017 to 2019.
Unless otherwise specified, analysis results from the On-Off background approach are always presented using the better matched Off List 1, with Off List 2 contributing to the evaluation of the systematic uncertainty only (see section \ref{sec:spectra} \& Table \ref{tab:spectra}).

\begin{table}
    \caption{Matching criteria of Off runs selected for the On-Off background analysis. } 
    \label{tab:matchoff}
    \centering
    \begin{tabular}{cccc}
     Dataset & Mean $\Delta\theta_z$ & Mean $\Delta t$ & Livetime \\
     \hline
     Off List 1 & $0.23^\circ\pm 0.04$ &$5.7\pm0.7$\,s &  17.9\,h \\ 
    Off List 2 & $0.97^\circ\pm0.03$ & $4.4\pm0.4$\,s &  20.8\,h \\ 
    \end{tabular}
    \tablefoot{
    The average discrepancy between On and Off runs in Zenith angle $\theta_z$ and run duration $t$ is indicated for each Off list. The resulting livetime after run quality selection is also quoted, with the difference in livetime due to tighter quality cuts and matching criteria used for Off List 1. We note that the two sets of Off runs are fully independent. 
    }
\end{table}

The second background estimation approach used is the field-of-view (FoV) method, which uses an acceptance model to estimate the expected level of background counts throughout the field of view \citep{BergeFunkHinton07}. Regions of significant $\gamma$-ray emission are excluded and the predicted number of background counts is normalised to the excess counts outside of these exclusion regions. 

\begin{figure}
    \centering
    \includegraphics[width=\columnwidth]{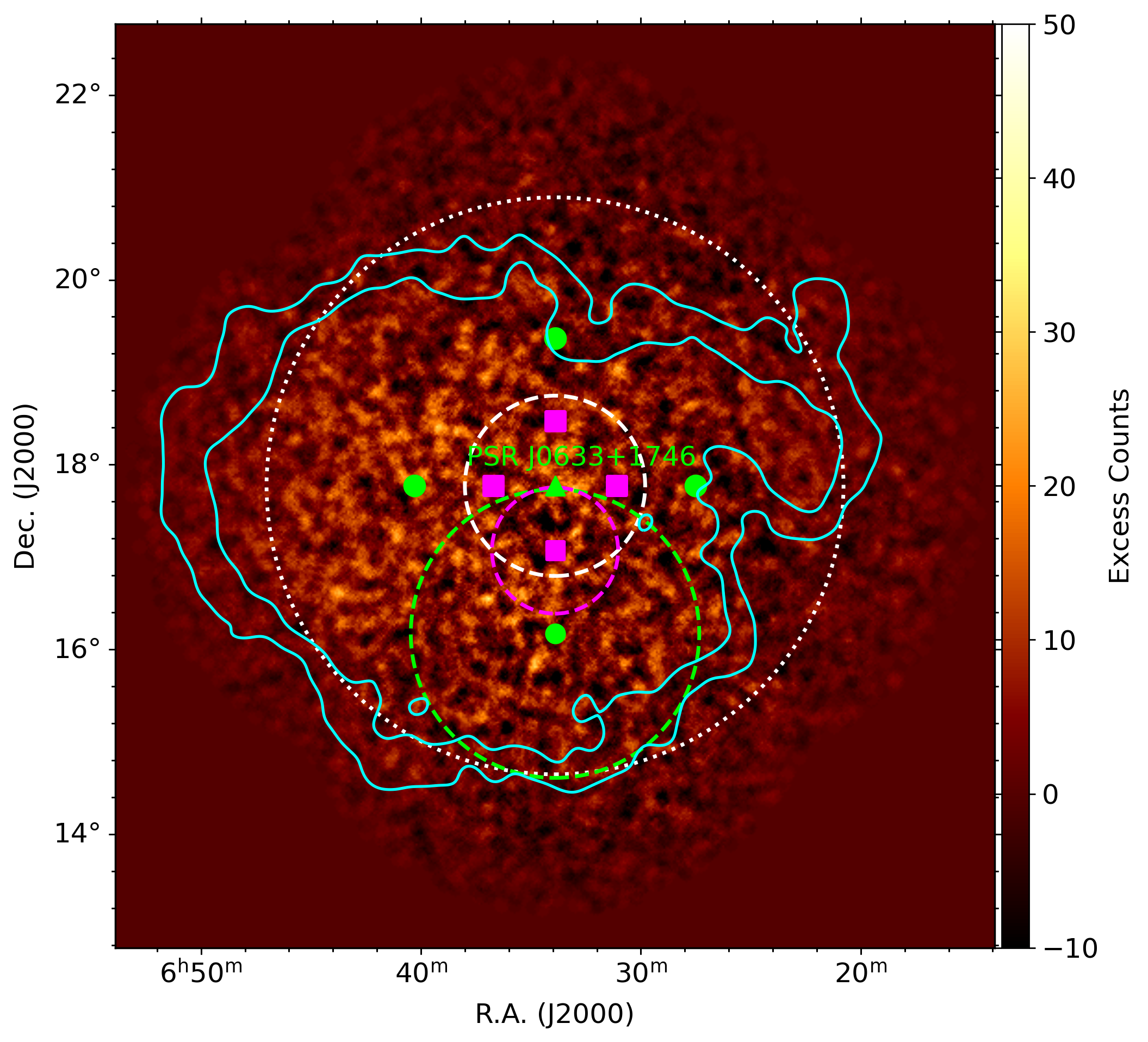}
    \caption{Observation positions corresponding to the 2006--2008 data set (magenta) and the 2019 dataset (green) at offsets of $0.7^\circ$ and $1.6^\circ$ from the pulsar respectively. The white dashed circle indicates the test On region with $1^\circ$ radius around the pulsar, whilst the white dotted circle indicates the $3.2^\circ$ radius region beyond which the background is normalised. 
    Green and magenta circles indicate the radius around an observation position with the same offset as the pulsar.  
    Counts map and contours are otherwise as in Fig. \ref{fig:maps}, left. }
    \label{fig:obspossketch}
\end{figure}

For both background estimation approaches, the background counts are normalised to the On counts in the region at radii $\theta >3.2^\circ$ away from the pulsar location (that is twice the offset of the observation positions), as shown in Fig. \ref{fig:hawc_onoff}. Excess counts maps constructed using both background estimation approaches are shown in Fig. \ref{fig:maps}, using a $0.08^\circ$ correlation radius. The observation positions used in 2006-2008 and in 2019 are indicated on the sky map shown in Fig. \ref{fig:obspossketch}. 

Some evidence for run-by-run variation in the Off count rate was investigated and found to be related to the atmospheric conditions, as indicated by the Cherenkov Transparency Coefficient, CTC \citep{HAHN201425}. The Pearson correlation coefficient for the Off count rate and CTC is $\sim0.7$ for both Off Lists. 
A correction for the variation in atmospheric conditions between the On and Off runs is implemented via two methods; using the ratio of the CTC between On and Off runs; and using the ratio of the number of hadron-like events in On and Off runs. Both methods are found to remove the correlation between the off count rate and the CTC. 
This correction is applied prior to the normalisation of On counts to radii $>3.2^\circ$. For the remainder of this paper only this dataset obtained in 2019 is used.

\subsection{Background estimation: Systematic tests of the On-Off and FoV methods}
\label{sec:syst}

As the analysis of sources with highly extended $\gamma$-ray emission is inherently challenging for IACTs, we performed a range of tests to assess the robustness of the analysis techniques and to quantify systematic uncertainties.
We performed an On-Off background analysis on two extragalactic regions with no expected $\gamma$-ray emission, using comparable parameters to the analysis of the Geminga region, that is with the same set of matching criteria for Off runs. For this purpose we used data obtained in 2017 and 2018 with the telescopes CT1-4 on two Dwarf Spheroidal galaxies, namely Reticulum II and Tucana III \citep{2020PhRvD.102f2001A_DES}. The average zenith angles of these datasets are $42.4^\circ$ and $39^\circ$ respectively, comparable to the $43.5^\circ$ average value of the Geminga dataset. No evidence for significant emission within the target region is found. The significance distributions for both dwarf spheroidals are found to be compatible with background, whereas the significance distribution for the Geminga region 
is asymmetric and exhibits a clear tail towards high significance. Also, the mean of the distribution is greater than zero for the Geminga region, reflecting the fact that the emission fills the field of view, whereas the distribution mean is compatible with zero for the test fields. 
This lack of signal from the test fields associated with dwarf spheroidal galaxies when using the same method provides confidence in the clear signal seen from the Geminga region. 
These empty fields are also analysed using the field-of-view background method; again, no significant emission is seen.

\begin{figure}
    \centering
    \includegraphics[width=\columnwidth]{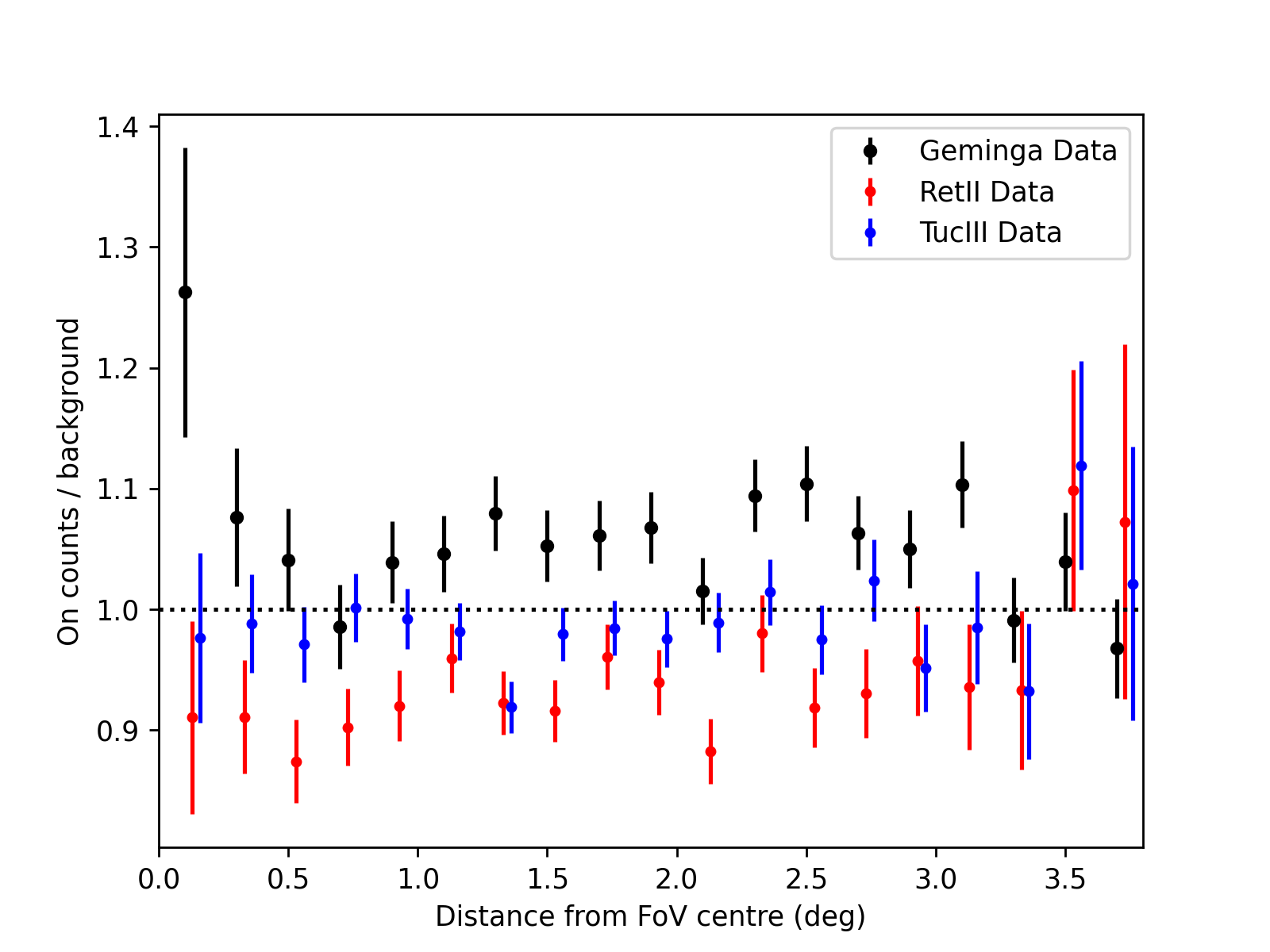}
    \caption{Ratio of On counts to background counts using the On-Off background method as a function of radial distance from the centre of the field of view. Shown are two empty sky regions taken on the dwarf spheroidal galaxies Reticulum II and Tucana III, as well as Geminga data. }
    \label{fig:empty_bg}
\end{figure}

Fig. \ref{fig:empty_bg} shows the ratio of On counts to background counts with distance from the FoV centre, with renormalisation to the background at a radial distance of $>3.2^\circ$ from the centre of the field of view. There are indications for a bias in the Ret\,II dataset with an On counts to background ratio lower than 1.0 out to beyond 3$^\circ$ (that is out to the radius which is used for normalisation to background). However, for the Tuc\,III dataset, the background normalisation results in approximately the expected level, averaging around 1.0. Correcting for a $\sim10\%$ bias for the region within $1^\circ$ radius of the pulsar would lead to a $\sim30\%$ increase in significance for the values quoted in Table \ref{tab:signif}. Given the limited number of tests performed, we conservatively estimate a systematic error of the order of $30\%$. A similar bias is found using the FoV background method, averaging $\sim6\%$ at radii $<1^\circ$. 

The field of view around the Geminga pulsar contains a bright star of magnitude 1.9, that leads to increased Night-Sky-Background (NSB) in the region. We investigated whether any dependence of the excess counts on the NSB in an On-Off analysis is seen, and found no correlation between the NSB rate and excess counts per run.

\section{Analysis results}
\label{sec:anaresults}
\subsection{Morphology of the $\gamma$-ray emission}
\label{sec:morphology}

\begin{figure}
    \centering
    \includegraphics[width=\columnwidth]{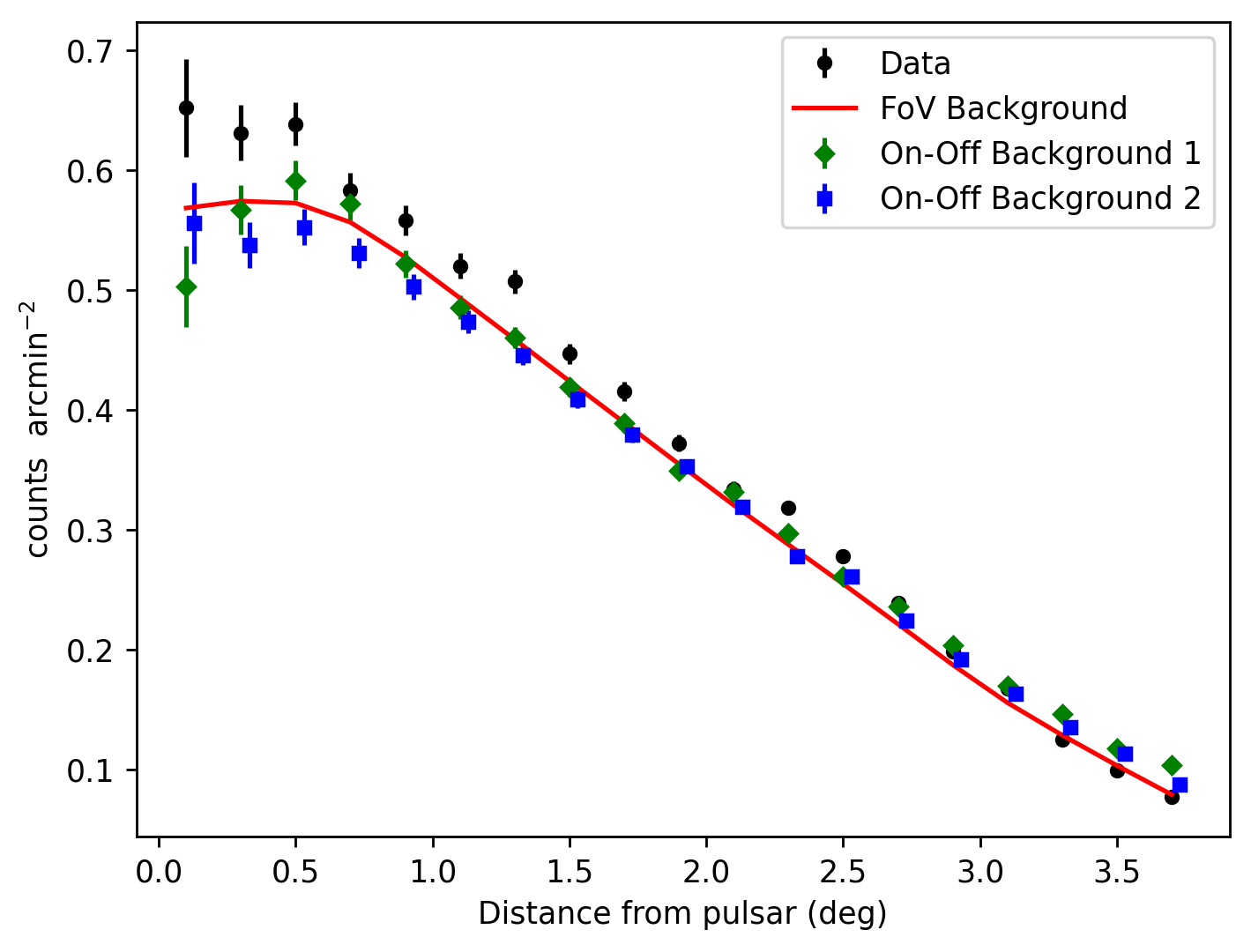}
    \caption{Radial profile without acceptance correction applied from uncorrelated maps, showing counts per unit area with radial distance from the pulsar at energies $>0.5$\,TeV, error bars are statistical. The different background methods show reasonable agreement.} 
    \label{fig:slices}
\end{figure}

Fig. \ref{fig:slices} shows the On data and background counts as a function of radial distance from the pulsar. The estimated background from the FoV and On-Off background methods (using two different Off run lists) are shown, and normalised to the data at radii $>3.2^\circ$, with a statistical uncertainty on the normalisation at the 1\,\% level. An excess of On data above background counts can be seen for all three estimations of the background level, particularly towards the innermost radii. The levels of background counts from the different methods are broadly consistent, with the remaining mild discrepancies providing an indication of the level of background systematics, estimated as $\leq 5\%$. Fig. \ref{fig:slices} provides further confidence that the detected emission is significant above the level of Galactic diffuse emission, which would peak towards the Galactic plane (the Geminga pulsar being situated towards the Galactic anti-centre at a galactic latitude of $4.27^\circ$).

The extension of the $\gamma$-ray emission is of particular relevance to studies of energetic particle transport in the region. Therefore, we tested the extent of the region of significant emission, firstly by varying the integration region radius (the radius around the pulsar within which the significance of the $\gamma$-ray emission is computed). Once the significant $\gamma$-ray emission is fully contained, the significance curve will start to flatten with increasing radius.

\begin{figure}
    \centering
    \includegraphics[width=\columnwidth]{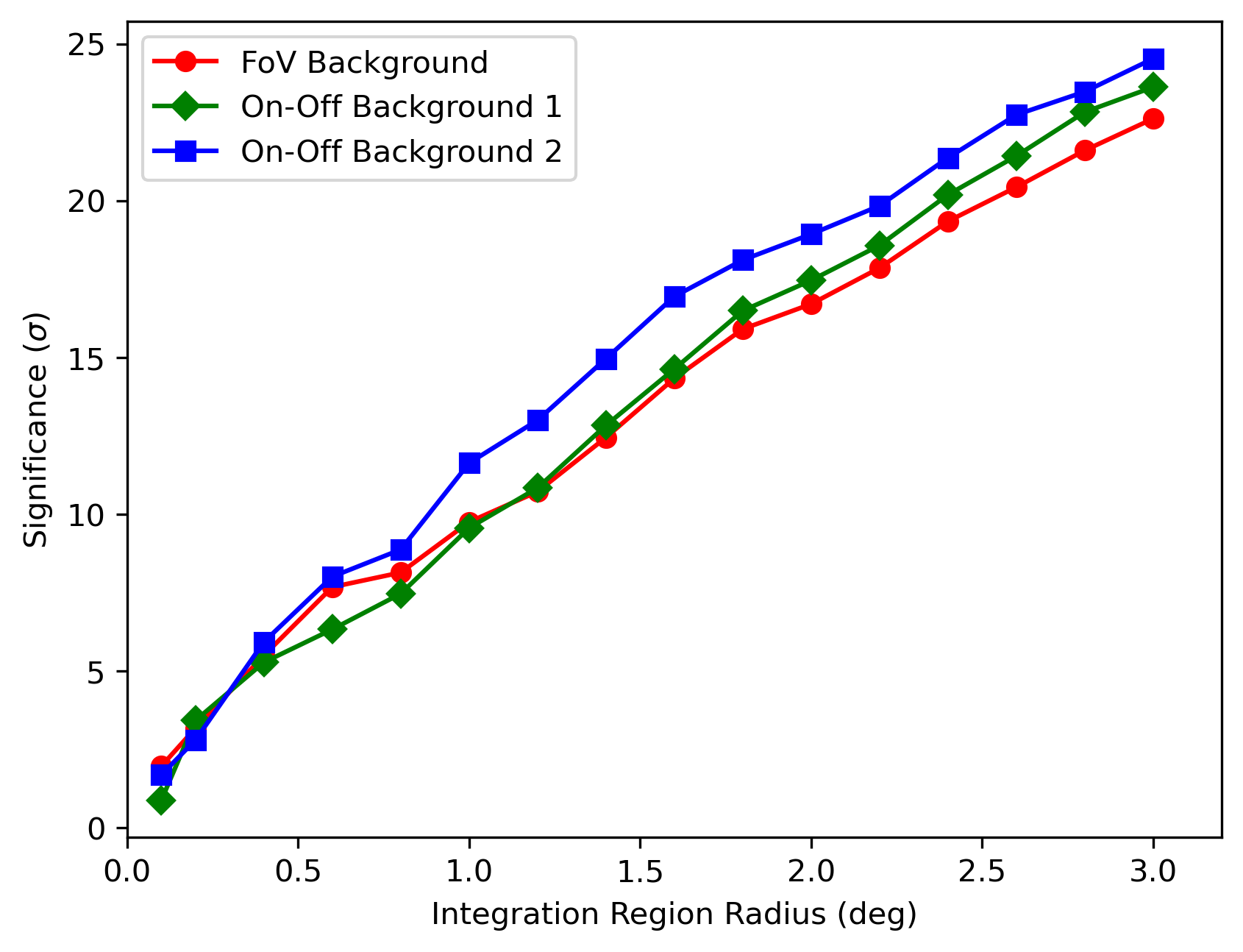}
    \caption{Significance with integration region radius around the pulsar for FoV and On-Off background estimation analyses. Emission within 1 degree radius of the pulsar is detected with a significance of $\gtrsim 8\sigma$ with all analyses.}
    \label{fig:intradsig}
\end{figure}

Fig. \ref{fig:intradsig} shows the significance with radius for the two independent background estimates. 
Notably, neither of the curves flatten within a radius out to $3^\circ$, indicating that the significant $\gamma$-ray emission is not yet fully contained by the integration region.
Secondly, the shape of the curves is consistent between the FoV and On-Off background methods; this is reassuring confirmation in the extent of significant emission provided by these fundamentally different approaches. 
Lastly, this analysis shows that the emission within a radius $<0.25^\circ$ around the pulsar is not significant.  
This further confirms that although concentrated towards the pulsar, the majority of the emission is spread out to larger radii. 

Although the analysis is limited by systematic uncertainties, we attempt to quantify the significance of a possible asymmetry as follows. Firstly, an acceptance-corrected azimuthal profile of the emission within a $3^\circ$ radius around the pulsar is constructed. 
The peak of the azimuth profile is found to be at $105^\circ\pm 3^\circ$ measured anticlockwise from the north, and is used to divide the region into two halves.  
From each semicircular half, radial surface brightness profiles of the emission, corrected for acceptance, are constructed in the two independent regions. 
Indications for a higher level of emission in one region compared to the other are seen at radii $>1^\circ$ (for example Fig. \ref{fig:maps}), however 
this is found to be neither statistically significant nor independently verified in the cross-check analysis.

To quantify whether the emission is significantly offset from the pulsar, despite the inability to measure the true extent, we evaluated the barycentre and 68\% containment radius of the excess counts, with acceptance correction applied. 
Indications for an offset of the emission centroid from the pulsar are found for all background methods, yet are nevertheless compatible with the pulsar location when systematic errors are taken into account. 
The emission centroid has an offset of $0.6^\circ$, at R.A. $99.1^\circ \pm 0.1^\circ \pm 0.5^\circ$ and Dec $17.7^\circ \pm 0.1^\circ \pm 0.5^\circ$ where errors are statistical and systematic (as estimated from the difference between the background methods). 
Evaluating the containment radius in three energy bands (0.5 - 2\,TeV, 2 - 8\,TeV, and 8 - 40\,TeV) no evidence for significant energy-dependent morphology is found. These containment radii in energy bands are summarised in Table \ref{tab:ebands_cont}. 

\begin{table}[h!]
    \caption{Containment radii at 68\%, $\theta_{68}$ of the excess emission, evaluated in different energy bands for both On-Off and FoV background methods. }
    \label{tab:ebands_cont}
    \centering
    \begin{tabular}{ccc}
    Energy Band & On-Off $\theta_{68}$($^\circ$) & FoV $\theta_{68}$($^\circ$) \\
    \hline
    0.5 - 2\,TeV & $2.7 \pm 0.2$ & $2.2 \pm 1.6$ \\
    2 - 8\,TeV & $1.9 \pm 0.8$ & $1.8 \pm 0.7$ \\
    8 - 40\,TeV & $2.5 \pm 1.2$ & $2.5 \pm 0.5$ \\
    0.5 - 40\,TeV & $2.4 \pm 0.1$ & $1.8 \pm 0.3$ \\
    \end{tabular}
    \tablefoot{The energy range of 8\,TeV - 40\,TeV is chosen to match the energy range of the analysis by the HAWC collaboration \citep{HAWC17}.}
\end{table}

Although TeV-bright $\gamma$-ray emission may be expected on angular scales $\lesssim0.1^\circ$ corresponding to the size of the X-ray PWN, with this analysis we see no indications of a separate component at these scales. Identifying such a separate emission component is challenging given the evolved state of the system, the proper motion of the pulsar (which is fast compared to the cooling time of electrons at the nearby distance of Geminga) and the predominance of inverse Compton emission from electrons already escaped from the PWN region.

\subsection{Spectral results}
\label{sec:spectra}

\begin{figure}
\includegraphics[width=\columnwidth]{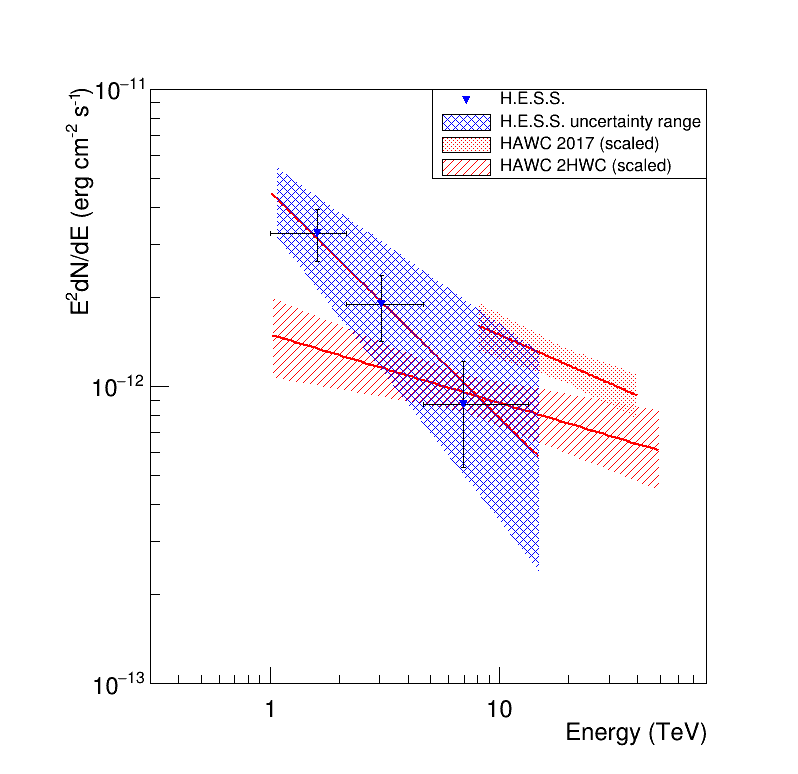}
\caption{Spectral energy distribution of the $\gamma$-ray emission within $1^\circ$ radius of the Geminga pulsar. Spectra from two HAWC analyses of Geminga are shown for comparison \citep{HAWC17,2HWC17}, with flux normalisation scaled to match the sub-region from which the H.E.S.S. spectrum is extracted. Error bands include systematic uncertainties for H.E.S.S. but are statistical only for HAWC. }
\label{fig:sed}
\end{figure}

We perform a spectral analysis for a circular region with 1$^{\circ}$ radius centred on the pulsar using the On-Off background approach; the resulting spectrum is shown in Fig. \ref{fig:sed} with parameter values for a best fit power law quoted in Table \ref{tab:spectra}. Due to the quality cuts applied for a spectral analysis, the energy threshold of the analysis is 1\,TeV. The spectrum obtained with HAWC for both the disc morphology of \cite{2HWC17} and the diffusion model of \cite{HAWC17} are also shown in Fig. \ref{fig:sed}, scaled for comparison to account for the different sizes of the regions probed. 
The scaling factor for the HAWC spectrum from \cite{2HWC17} corresponds to the ratio of the disc areas for a $2^\circ$ radius and a $1^\circ$ radius, the region probed by this H.E.S.S. analysis. For the HAWC spectrum from \cite{HAWC17}, we rescale according to the ratio between the diffusion model integrated out to 1$^\circ$ radius and the full integral, assuming that the spectral index is not radially dependent.

\begin{table}
\caption{Fit parameters to spectra obtained from a $1^\circ$ radius region around the pulsar using the two Off Lists.} 
\label{tab:spectra}
\centering
\begin{tabular}{cccc}
Dataset & Index & $\phi_0$ (cm$^{-2}$s$^{-1}$TeV$^{-1}$) & Signif. \\
\hline
Off List 1 & $2.76 \pm 0.22$ & $(2.81\pm0.71)\times10^{-12}$ & 6.7 $\sigma$\\
Off List 2 & $2.62 \pm 0.14$ & $(3.55\pm0.68)\times10^{-12}$ & 8.9 $\sigma$
\end{tabular}
\tablefoot{ A power law spectral model $\frac{\mathrm{d}N}{\mathrm{d}E}=\phi_0\left(\frac{E}{E_0}\right)^{-\Gamma}$ is assumed, with index $\Gamma$ and  $E_0=1$\,TeV. The systematic errors are estimated at $\pm0.14$ on the spectral index and $\pm0.7 \times 10^{-12}$cm$^{-2}$s$^{-1}$TeV$^{-1}$ on the flux normalisation for both Off Lists. }
\end{table}

Fig. \ref{fig:sed} shows that the scaled HAWC flux and the H.E.S.S. measurement are consistent at energies $\gtrsim 5$\,TeV and compatible within the systematic errors of both measurements \citep{HAWC17,2HWC17}. From \cite{HAWC17}, the best fit spectral measurement has a spectral index of $2.34 \pm 0.07$ and flux normalisation of $13.6^{+2.0}_{-1.7}\times 10^{-15}$  TeV$^{-1}$cm$^{-2}$s$^{-1}$ at 20\,TeV. We use this latter spectrum obtained from a dedicated analysis of HAWC data on the Geminga region for modelling purposes. The systematic errors of the H.E.S.S. spectral measurement are $\sim0.2$ on the index and $\sim30\%$ on the flux, obtained from the cross-check between the background models. The $\sim 8$\% bias seen at radii $<1.0^\circ$ in Fig. \ref{fig:empty_bg} corresponds to a $\sim30\%$ systematic on the flux, assuming these two contributions are uncorrelated, the total systematic error on the flux is of the order of $40\%$. The systematic errors of the HAWC spectral measurement are $0.2$ on the index and $50\%$ on the flux.
We note that the measured HAWC flux is dependent on the morphological model assumed for the emission - here we show the disc model, the approach most similar to the H.E.S.S. spectral measurement, which is also approximately a disc with $1^\circ$ radius.  

\subsection{Radial profiles}

\begin{figure}
\includegraphics[width=\columnwidth]{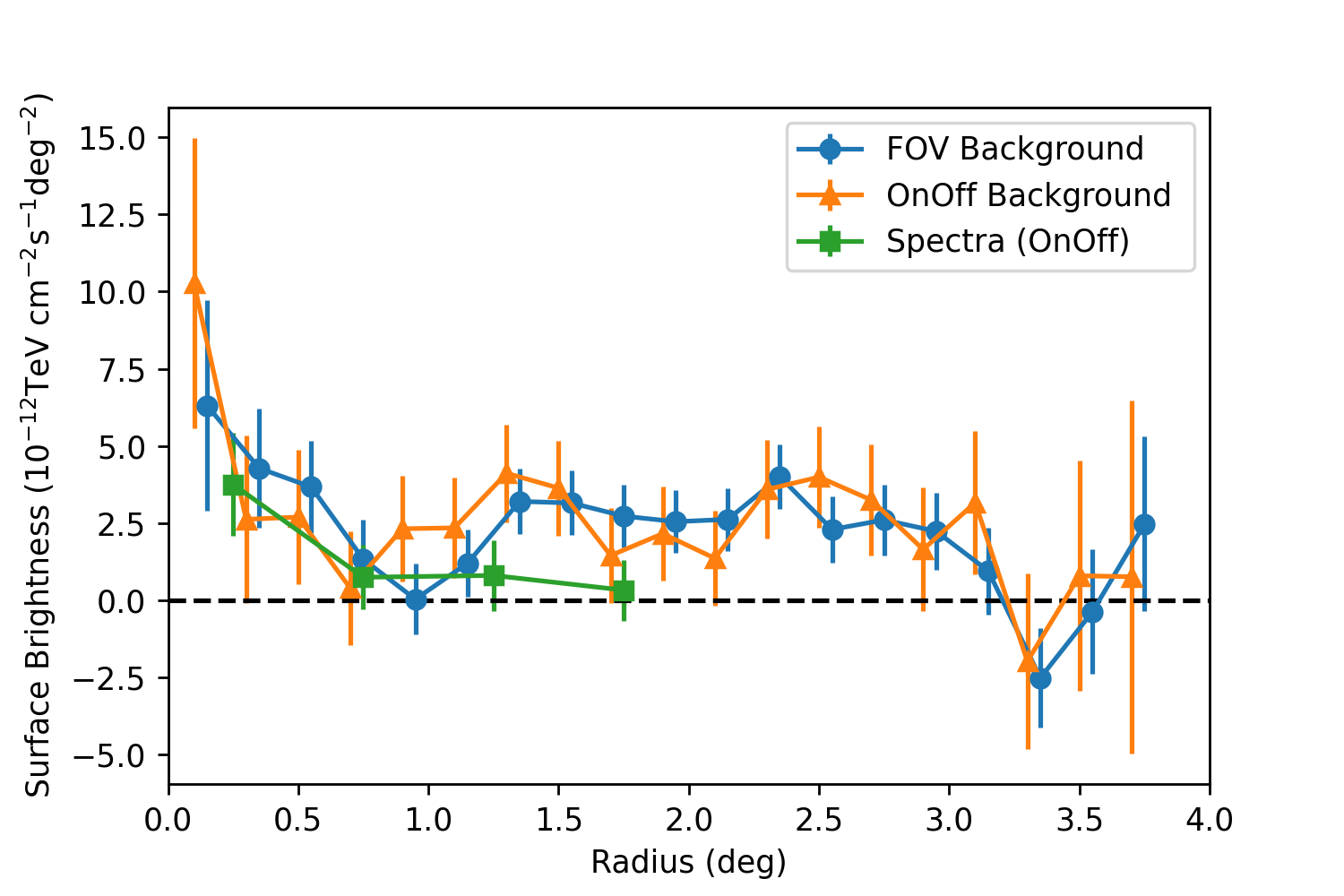}
\caption{Radial profile constructed using three independent methods; from $\gamma$-ray maps with FoV and On-Off background methods (Off list 1), and from a spectral analysis in consecutive ring regions around the pulsar. The latter is limited to radii $<2^\circ$ from the pulsar and is shown without an additional normalisation to background at radii $>3.2^\circ$ applied. }
\label{fig:profile}
\end{figure}

Acceptance-corrected radial profiles of the excess emission are constructed from the sky maps, using a circular region centred on the pulsar extending out to $\sim3.5^\circ$ radius.
The surface brightness profiles are constructed from acceptance-corrected excess maps with units of cm$^{-2}$s$^{-1}$deg$^{-2}$. 
For the conversion to surface brightness in units of TeV\,cm$^{-2}$s$^{-1}$deg$^{-2}$, a power law spectral model is used with index $\Gamma=2.76$ (corresponding to the best-fit spectral parameters, see Table \ref{tab:spectra}).
These profiles are presented in Fig. \ref{fig:profile}.

Using the On-Off background method, spectral analyses are additionally performed in concentric rings of $0.5^\circ$ width, out to a radius of $2^\circ$. 
The data is fit in each ring using a power law spectral model, from which radial profiles of the emission are constructed. The best fit model is integrated within the energy range $0.5-40$\,TeV to obtain the flux as a function of radius within a specific energy band. This radial profile is also shown in Fig. \ref{fig:profile}. The resulting radial surface brightness profiles are therefore found to be compatible between the normalised profiles from maps constructed with On-Off background and from maps constructed with the FoV background. The radial surface brightness profile from a spectral analysis is however limited to radii $<2^\circ$. In order to normalise this radial profile constructed from spectra, we applied the same normalisation as that obtained for the radial profile for On-Off background constructed from sky maps.

These radial profiles also show the normalisation to background at radii $>3.2^\circ$, which implies that the flux is likely underestimated over the region.
The innermost radial bin corresponds to a region $0.2^\circ$ around the pulsar, comparable in size to that of the region probed by X-ray instruments. A flux measurement is obtained from the surface brightness within this radial bin, assuming that the spectral index does not vary with radius. 
In section \ref{sec:model} these radial profiles, together with the spectral energy distribution, are fit with a diffusion model.

\section{Modelling}
\label{sec:model}

\subsection{Diffusion model description}

For the particle transport we consider energy-only dependent diffusion, energy losses, and a continuous point-like source of electrons:

\begin{eqnarray}
    \partial_t N(E,\vec{r},t) - D(E)\Delta N(E,\vec{r},t) + \partial_E[b(E)N(E,\vec{r},t)] \nonumber \\ = Q(E,t)\,\delta(\vec{r}-\vec{r_s}),
\end{eqnarray}
with $N(E,\vec{r},t)$ the density of electrons per energy, $D(E)$ the spatial diffusion coefficient, $b(E)$ the energy loss rate, and $Q(E,t)\delta(\vec{r}-\vec{r_s})$ the source term with $Q(E,t)$ the rate of produced electrons per energy and $\vec{r_s}$ the source position. The assumption of a point-like source of injection is motivated by X-ray observations of the Geminga pulsar wind nebula \citep{Pavlov10} showing emission up to $0.06$\,pc whereas the observed TeV emission extends up to $14$\,pc. An advection term is not included in this expression for the particle transport, as this would generate asymmetries to the $\gamma$-ray emission that are not seen in this analysis.

\subsubsection{Diffusion coefficient}
The diffusion coefficient $D(E)=D_0(E/E_0)^\delta$ is assumed to follow a power law, with index $\delta$ in the range $\delta \in [0.3,1]$, to cover the scale between between Kolmogorov turbulence \citep{Kolmogorov10.2307/51980}, Kraichnan turbulence \citep{Kraichnan}, and Bohmian diffusion \citep{BohmDiff}. For this model, $E_0 = 100\,$TeV as in \cite{HAWC17} is chosen to have a direct comparison of the diffusion coefficient with the HAWC results, whilst $D_0$ is kept as a free parameter of the fit.  

\subsubsection{Energy losses}
\label{section_energy_losses}

Synchrotron and Inverse Compton energy losses are considered. We assume the presence of three different photon fields; the Cosmic Microwave Background, Infra-Red from dust, and Optical radiation from star light (see also appendix \ref{sec:EnergyLosses}). 
Magnetic field values between $1$\,$\mu$G and $5$\,$\mu$G are tested, which correspond to inverse-Compton dominant and synchrotron dominant scenarios respectively.

\begin{figure}
    \centering
    \includegraphics[width=\columnwidth]{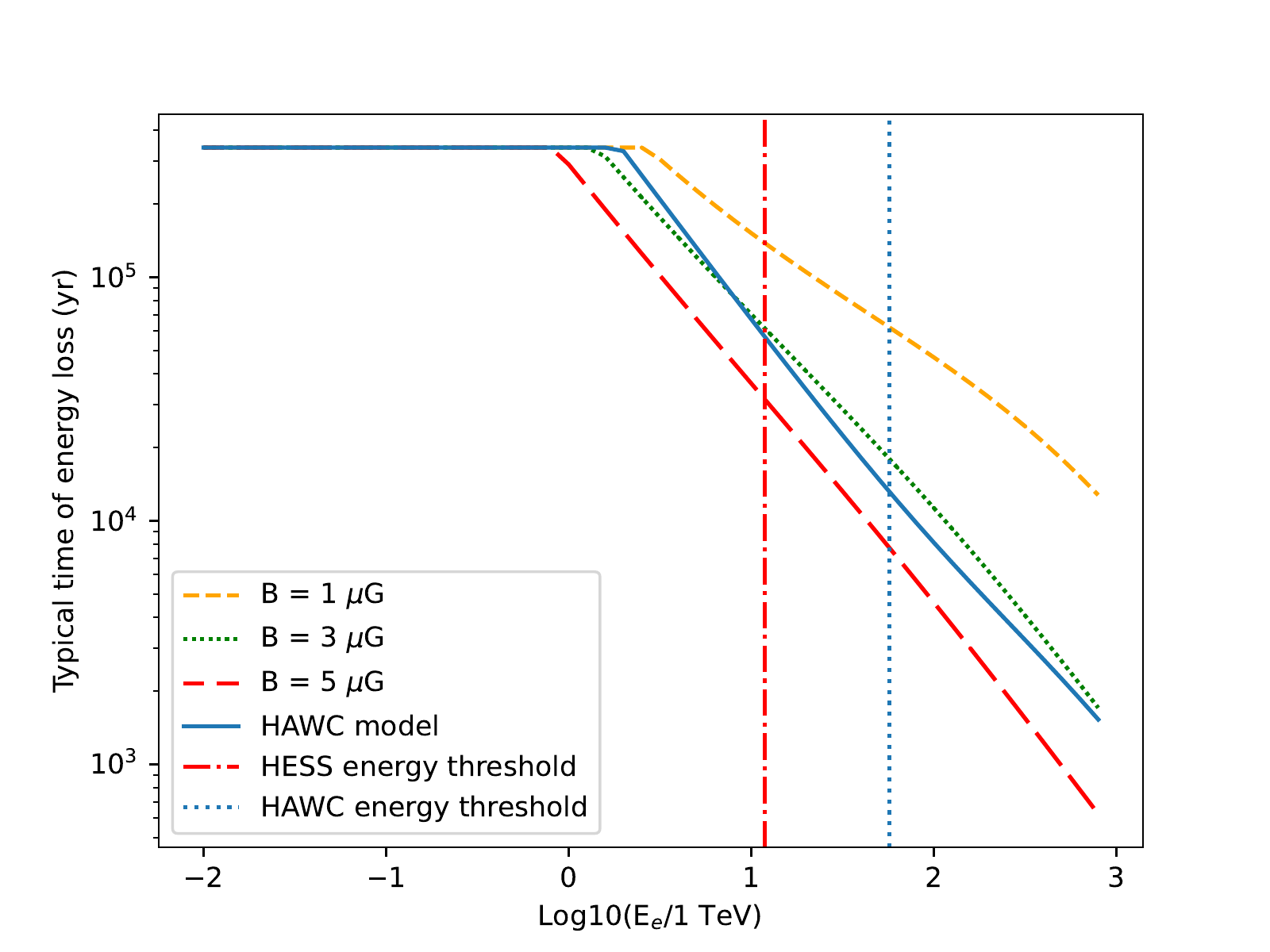}
    \caption{Typical energy loss time versus electron energy. If the particle loss timescale is longer than the pulsar age (here chosen as $340$ kyr), then the typical time $t_E(E)$ (see equation \ref{equ:typicaltime}) is taken as the pulsar age. The difference to the HAWC model is due to the approximation of energy losses made in the HAWC model.}
    \label{fig:loss_time}
\end{figure}

\noindent The typical time of energy losses is defined as: 
\begin{equation}
    t_E(E) = - \int_E^{+\infty} \frac{1}{b(E^\prime)} \mathrm{d}E^\prime~,
\label{equ:typicaltime}
\end{equation}
where $b(E^\prime)$ represents the total energy-dependent energy losses.
Eq. \eqref{equ:typicaltime} can be defined as the maximum age of an electron given its current energy. This simple calculation enables us to assess the age of the particles that can be observed, as shown in Fig. \ref{fig:loss_time}. Given the H.E.S.S. energy threshold converted to electron energy, the maximal age of particles is between $140$\,kyr for the $1$\,$\mu$G case and $32$\,kyr for the $5$\,$\mu$G case.

\subsubsection{Source term}

The total energy released by the pulsar since its birth, $W_0$, is defined as: 

\begin{equation}
    W_0 = \int^{T_*}_0 L(t) \mathrm{d}t~,
\end{equation}
where $T_*$ is the current pulsar age and $L$ is the pulsar luminosity function (see appendix \ref{appendix:pulsar}). We assume that the source term $Q(E,t)$ (defined as the rate of electrons per energy produced by the pulsar) follows a power law with an exponential cut off, and its normalisation follows the same time-dependence as the pulsar luminosity: 

\begin{equation}
    Q(E,t) = Q_0 (1+t/\tau_0)^{-(n+1)/(n-1)} (E/E_0)^{-\alpha}\exp{(-E/E_c)}~,
\end{equation}
where $\tau_0$ is the initial spin-down timescale, $n$ is the braking index, and $E_c$ the energy cut-off of electron spectra (see appendix \ref{appendix:pulsar} for the definitions of $\tau_0$ and $E_c$). For simplicity, $E_c$ is assumed to be constant during the lifetime of the pulsar. The normalisation of the source term can be related to the total energy released by the pulsar by: 

\begin{equation}
    \int^{+\infty}_{E_{min}} \int^{T_*}_0 Q(E,t) E \mathrm{d}E \mathrm{d}t = \eta W_0~,
\end{equation}
where the efficiency parameter $\eta$ is the proportion of the pulsar spin-down energy converted to electron-positron pairs that escaped from the close source environment to the extended emission region.

\subsection{Solution}

The solution to the diffusion energy loss equation as described in \cite{diMauroFermi2019PhRvD.100l3015D} is:

\begin{eqnarray}
    N(E,\vec{r},t) &=& \int^t_0 \mathrm{d}t_0 \frac{b(E_s(E,t_0,t))}{b(E)} \frac{1}{(\pi \lambda^2(t_0,t,E))^{3/2}} \nonumber \\ 
    &\times & \exp{\left(-\frac{|\vec{r}-\vec{r_s}|^2}{\lambda^2(t_0,t,E)}\right)}Q(E_s(E,t_0,t),t_0),
\end{eqnarray}
where $E_s(E,t_0,t)$ is defined as the energy of an electron that was produced at time $t_0$ and observed at time $t$ with energy $E$. This value is obtained by solving numerically: 

\begin{equation}
    t-t_0 = \int_E^{E_{s}} \frac{\mathrm{d}E^{\prime}}{b(E^{\prime})}~,
\end{equation}
while the diffusion radius $\lambda(t_0,t,E)$ is defined as: 

\begin{equation}
    \lambda^2 = 4\int_E^{E_{s}} \mathrm{d}E^{\prime} \frac{D(E^{\prime})}{b(E^{\prime})}~.
\end{equation}
Since the data we are using to constrain this model are radial profiles, we compute the mean density over the angle between $\vec{r}$ and $\vec{r_s}$ of electrons observed today at a given energy and distance from the current pulsar position $r$ can be written as:

\begin{eqnarray}
    N(E,r,T_*) &=& \int^{T_*}_0 \mathrm{d}t_0 \frac{b(E_s(E,t_0,T_*))}{b(E)} \frac{1}{(\pi \lambda^2(t_0,T_*,E))^{3/2}} \nonumber \\
    &\times & \exp{\left(-\frac{r^2 + r_s^2(t_0)}{\lambda^2(t_0,T_*,E)}\right)}Q(E_s(E,t_0,T_*),t_0)~,
\end{eqnarray}
with $T_*$ being the current pulsar age, where $E_s(E,t_0,T_*)$ is defined as the energy of an electron that was produced at time $t_0$ and observed at time $T_*$ with energy $E$. Since $r_s(t)$ is varying as a function of time, the impact of the asymmetric morphology is taken into account in our radial profiles. In addition to the source parameters, the density of electrons depends on the diffusion parameters ($D_0$ and $\delta$) and electron cooling parameters ($B$ field and IC photon field energy density and temperature). The parameter $r_s$ allows us to introduce the effect of the pulsar proper motion, and is defined as: $r_s(t) = V_T (T_* - t)$ where $V_T$ is the transverse speed of the pulsar.

\begin{figure}
    \includegraphics[ clip,width=\columnwidth]{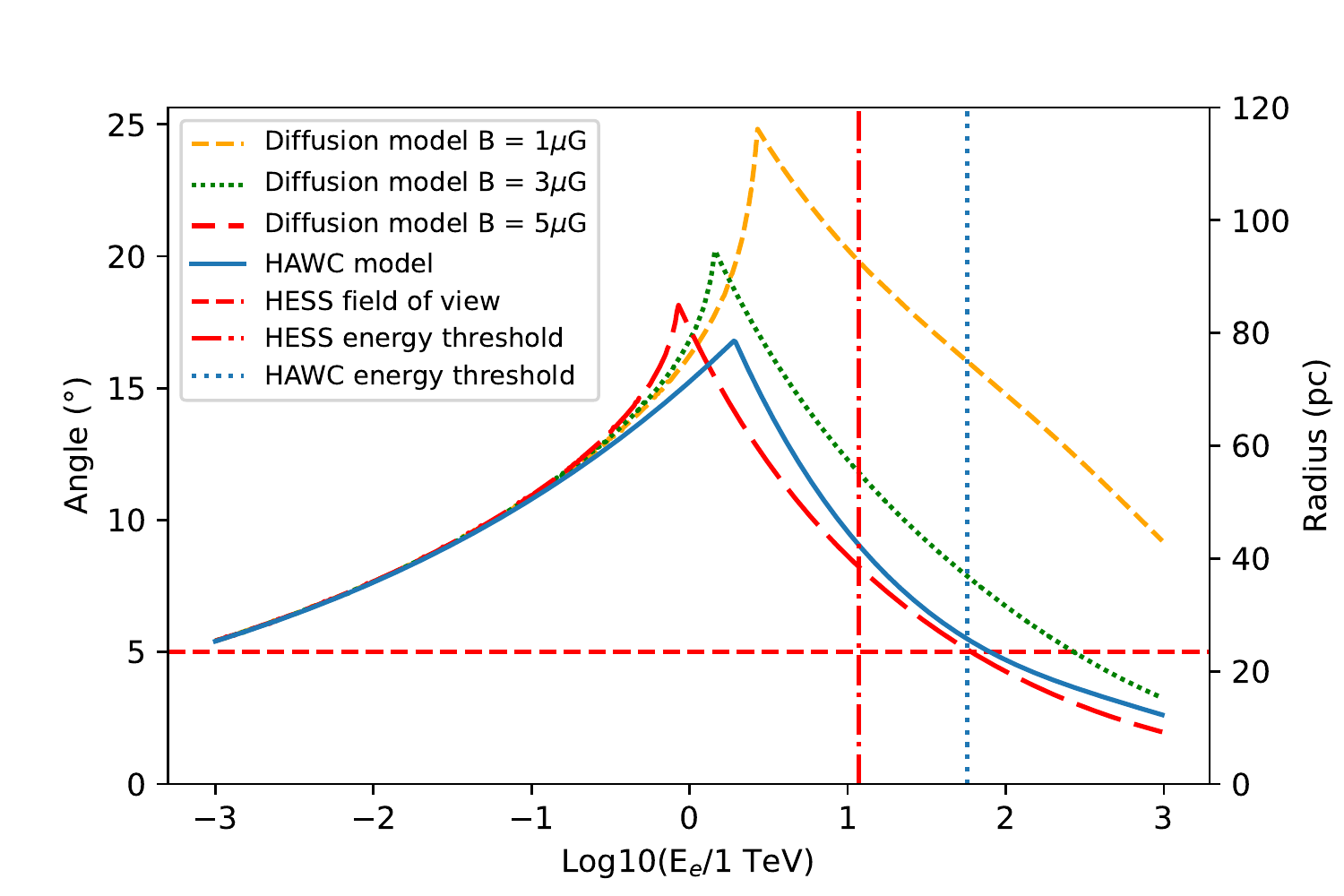}
    \caption{Comparison of the predicted typical diffusion scale $\lambda$ as a function of electron energy for different diffusion models and magnetic fields. The HAWC model is taken from \cite{HAWC17}. It is clear that the emission is predicted to exceed the H.E.S.S. field of view over most of the H.E.S.S. energy range (here converted to electron energy). }
    \label{fig:model_profile}
\end{figure}

As shown in Fig. \ref{fig:model_profile}, the diffusion radius peaks at the energy where the cooling time for electrons is equal to the pulsar age. For the Geminga pulsar with a characteristic age of $T_c = 342$\,kyr, this occurs at $\sim1$\,TeV. At lower energies, the electrons are uncooled and the diffusion radius $\lambda$ is limited by the diffusion parameters; whereas at higher energies, the diffusion radius is limited by the cooling time for electrons in the ambient magnetic field. 
Fig. \ref{fig:model_profile} demonstrates that the diffusion radius in the energy range of H.E.S.S. is much larger than the H.E.S.S. field of view, such that the full extent of the emission cannot be measured.

\subsection{Model joint fit with H.E.S.S., HAWC and XMM-Newton data}

The solution of the diffusion model provides us with the density of electrons per energy interval. Two more steps are needed to compare the diffusion model with the multi-wavelength data: 
Firstly, the electron density needs to be projected onto a 2D plane and converted to a density versus angle based on the pulsar distance. Secondly, the obtained electron density profile is then converted to the $\gamma$-ray flux through Inverse Compton (IC) scattering. This step makes use of the Naima package  \citep{naima}, with a photon field for IC approximated as three black-body components defined in appendix \ref{sec:EnergyLosses}.

The diffusion model described above is simultaneously fit to radial profiles and spectral energy distribution (SED) using a chi-square minimisation. The normalisation of the diffusion coefficient $D_0$ is used as a free parameter while the energy cut-off of the electron spectrum $E_{c}$ is tested as a free or fixed parameter in order to evaluate its significance in the final result. We performed this analysis using the combination of H.E.S.S. (radial profile and SED within $1^\circ$) and HAWC 2017 (radial profile) data (see section \ref{sec:spectra}). The HAWC SED from \cite{HAWC17} is not used in the fit but is superimposed to the model SED obtained with an integration radius of $10^\circ$. 
The GeV part of the Geminga halo is assumed to be highly offset compared to the actual pulsar position due to the proper motion of the pulsar. A detection of such emission is claimed by \cite{diMauroFermi2019PhRvD.100l3015D}, where a template fit, including the effect of pulsar proper motion, enabled the detection of emission between $10$ GeV and $100$ GeV 
with an offset position and extent consistent with that expected from the diffusion model fit to HAWC data. 

We decided not to include these data because our single zone model is not valid on the full extent of the claimed Fermi-LAT emission. Indeed, a single zone of diffusion for the Fermi-LAT extent scale cannot take into account the positron excess of AMS-02 as explained in \cite{2022arXiv220611803M}. 
A two zone diffusion model is beyond the scope of this paper, since the emission measured with H.E.S.S. is not extended enough to constrain such a model. In order to constrain the synchrotron contribution to the SED, we compared keV upper limits to the model SED, derived using XMM-Newton data in a $10'$ region around the Geminga pulsar as described in \cite{2019ApJ...875..149L}.

In order to explore the phase space whilst keeping $D_0$ and $E_{c}$ free in the fit, we conduct a parameter scan over five variables ($n,\eta,\alpha,B,\delta$), because a global minimisation procedure is found to be degenerate. These scanned parameters are listed in Table \ref{Table:Param}. All possible combinations of parameters are tested, leading to 243 different combinations to explore the full parameter space. Values for the parameters characterising the Geminga pulsar are taken from the ATNF pulsar catalogue \cite{Manchester05}. The initial period of the pulsar is assumed to be $15$\,ms since its value is not affecting results above the H.E.S.S. energy threshold. 
Two of the free parameters are further constrained by validity conditions: $P_0 < P_*$ and $\eta < 1$.

\begin{table}[h!]
\caption{Input parameters for the diffusion model. }
\begin{center}
\begin{tabular}{ ccc } 
 Parameter & Description & Value(s) \\ 
 \hline
 $L_*$ & Current Luminosity & $3.2 \times 10^{34}$ erg/s \\ 
 $T_c$ & Pulsar age & $342$ kyr \\ 
 $n$ & Braking index & [$1.5$, $3$, $4.5$]\\
 $P_0$ & Initial period & $15$ ms \\
 $P_*$ & Current period & $0.237$ s \\
 $d$ & Distance & $250$ pc\\
 $V_T$ & Transverse velocity & $211$ km/s \\
 $\eta$ & Electron efficiency & [$0.01$, $0.1$, $0.5$] \\ 
 $\alpha$ & index of electron & [$1.8$, $2.0$, $2.2$] \\
  & injection distribution &  \\
 $E_c$ & Energy cut off of electron  & [free,$1$ PeV] \\
  & particle distribution &  \\
 $B$ & Ambient magnetic field & [$1$, $3$, $5$]  $\mu$G \\ 
 $D_0$ & Normalisation & free  \\ 
 & of diffusion coefficient  &   \\ 
 $\delta$ & power law index   & [$0.3$, $0.6$, $1$]\\
  & of diffusion coefficient  & \\
\end{tabular}
\end{center}
\tablefoot{$D_0$ is the normalisation at an electron energy $E_0=100$\,TeV. }
\label{Table:Param}
\end{table}

The fits are performed for the results using both On-Off background lists separately, as well as for the main and cross-check analyses. For each fit result, the p-value is computed and is used to define the goodness of fit. A p-value $> 0.003$, corresponding to three standard deviations is taken, as a threshold to define a good fit. This can be considered as a fairly relaxed threshold since the systematic uncertainties of both H.E.S.S. and HAWC data are important. 

\begin{figure}
\centering 
\includegraphics[width=\columnwidth]{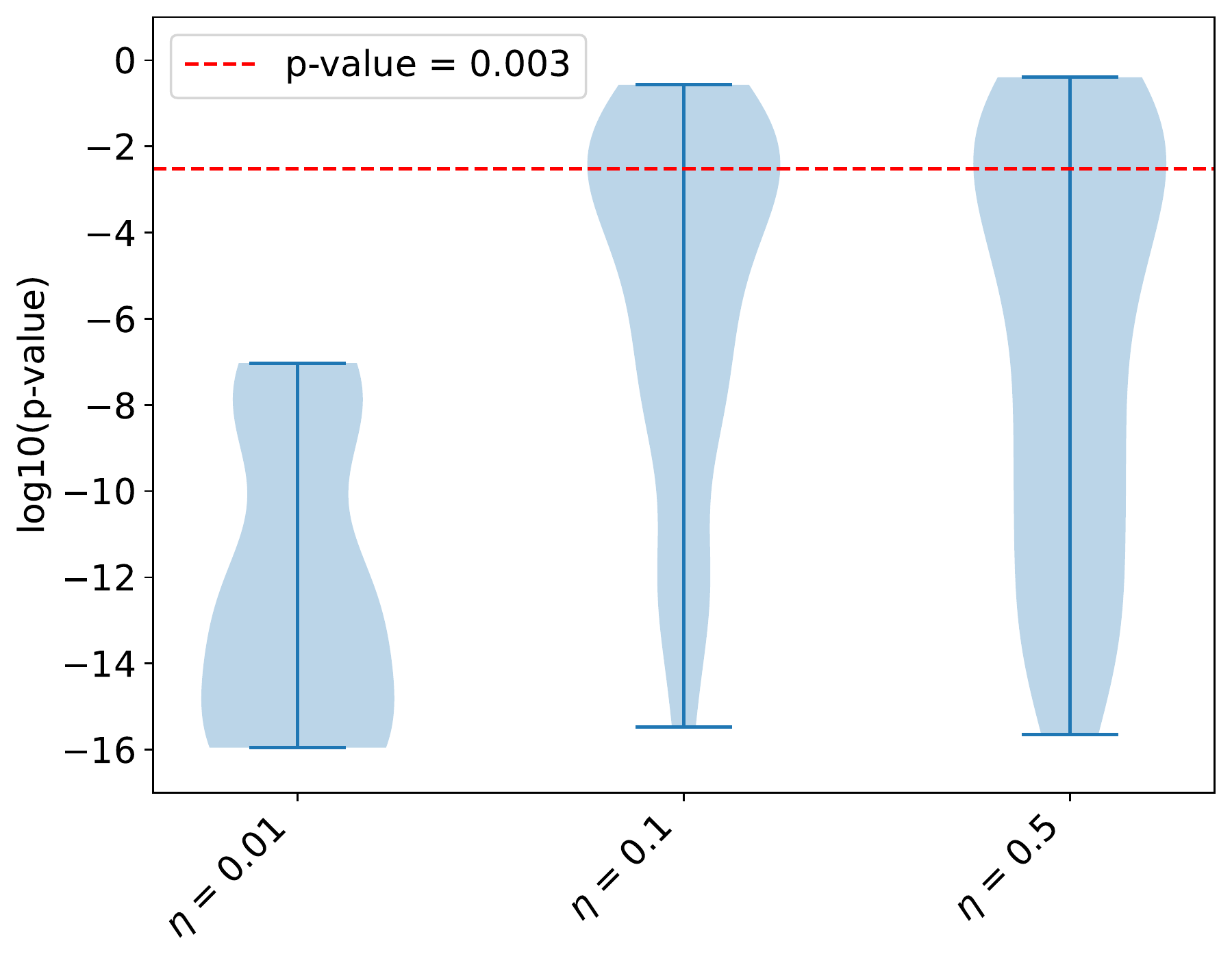}

\caption{Distribution of p-value for the scanned parameter combinations presented in the table \ref{Table:Param}, with different values of $\eta$ parameter, using the H.E.S.S. and HAWC datasets in combination with $E_c$ fixed to 1\,PeV. This distribution is presented as violin plot, illustrating kernel probability density. The width of the shaded area represents the proportion of the number of fits located at these position. The upper and lower bar represent the maximum and minimum values. Models with p-value $<0.003$ are excluded.}

\label{fig:fitResults_etap}
\end{figure}

The p-value distribution as a function of the electron efficiency from the fixed $E_c$ scan is presented in Fig. \ref{fig:fitResults_etap}, where only those good fit models with p-value $>0.003$ are retained. The combination of H.E.S.S. and HAWC data excludes an efficiency of injection and electron-positron pair conversion lower or equal to $1$\,\% at the $5\,\sigma$ level. However, no significant preference is found between the values scanned for the power law index of the diffusion coefficient $\delta$, the ambient magnetic field $B$, injection index $\alpha$ and braking index $n$.

\begin{figure}
\centering 
\includegraphics[width=\columnwidth]{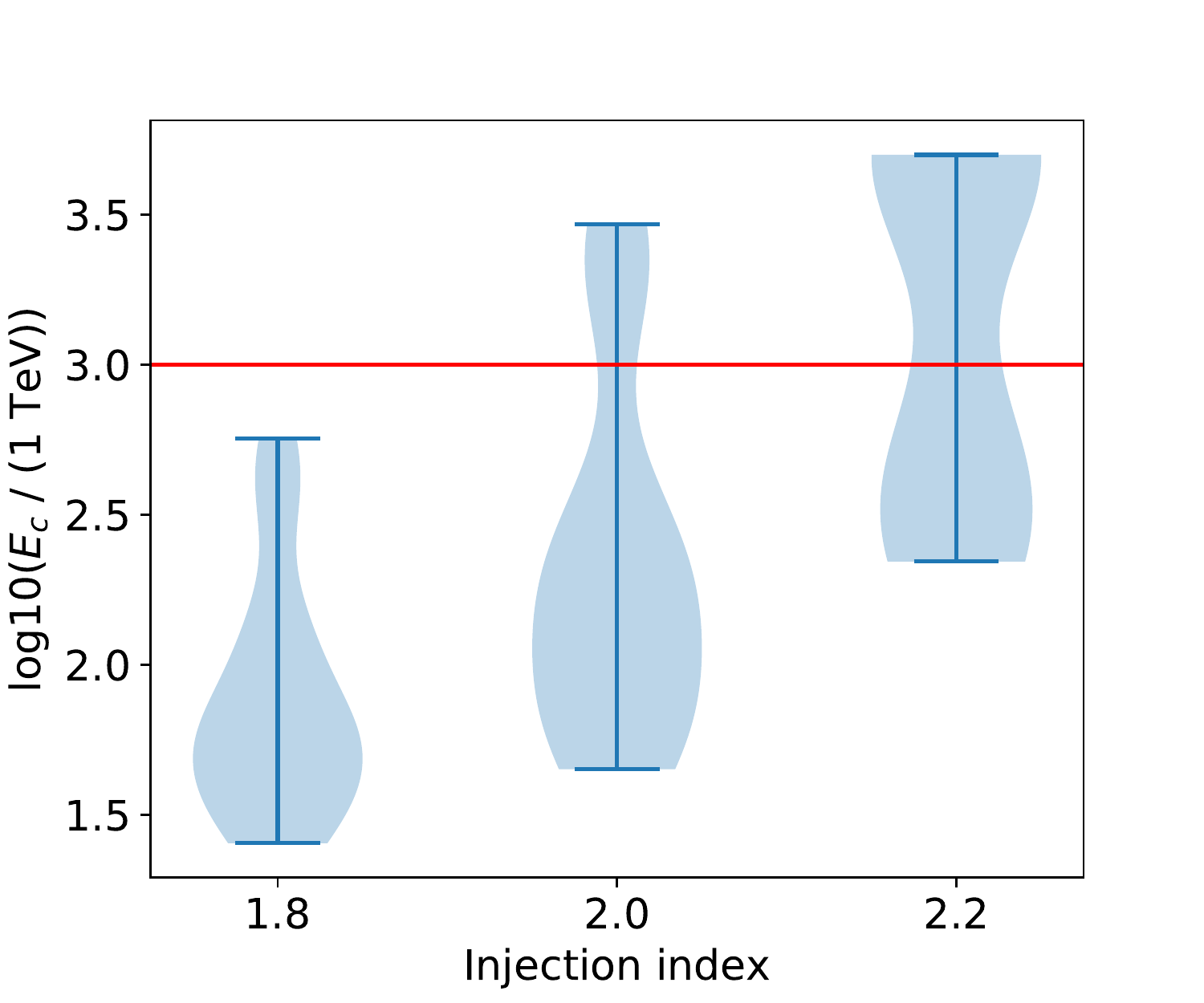}
\includegraphics[width=\columnwidth]{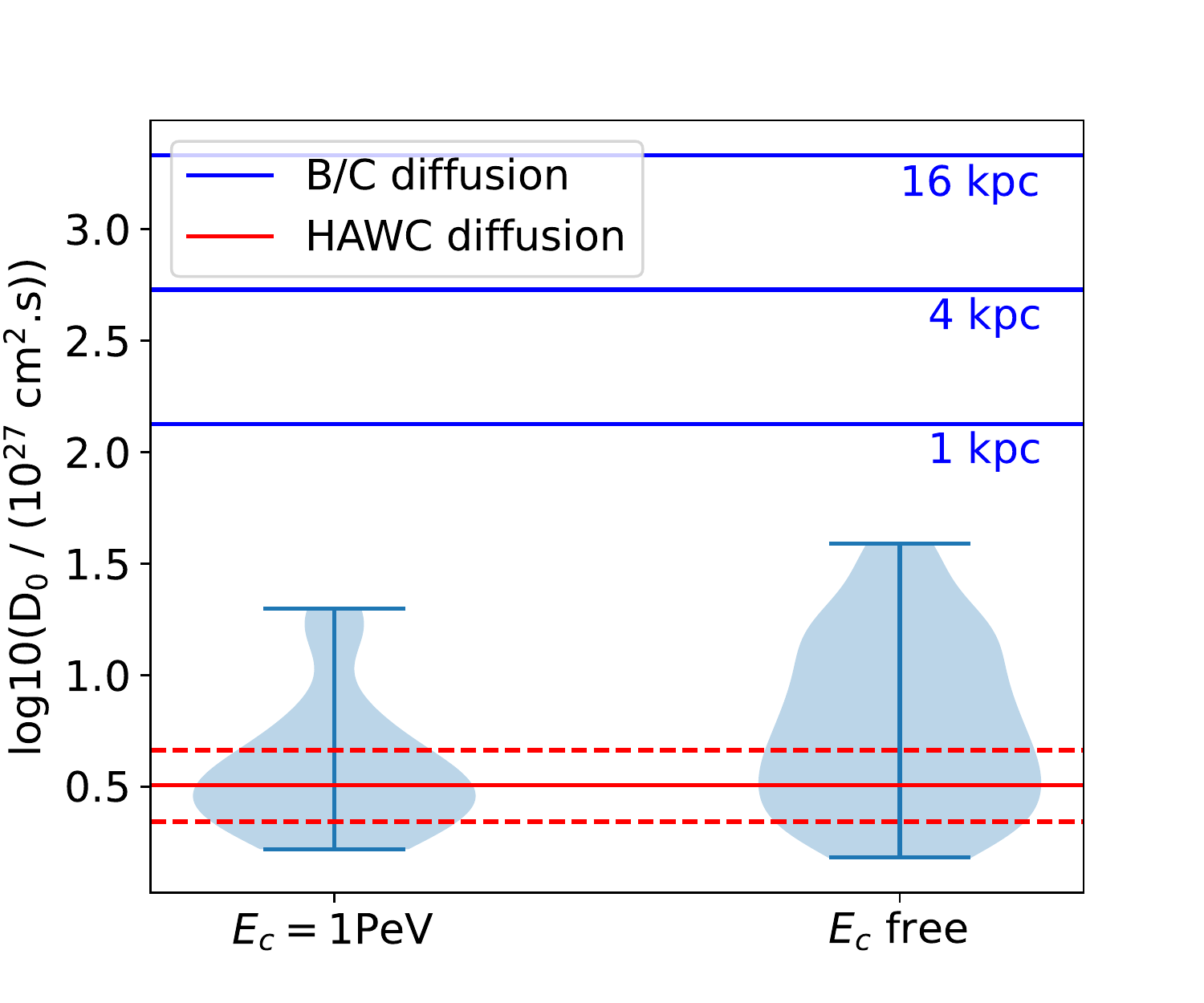}

\caption{Distribution of models with p-value $>0.003$, presented as a violin plot, illustrating the kernel probability density. The width of the shaded area represents the proportion of number of fits located at these positions. The upper and lower bars represent the maximum and minimum values.
Top: Fitted energy cut-off distribution as a function of injection index for the combined H.E.S.S. and HAWC dataset. The red horizontal line shows the value of energy cut-off chosen for the fixed $E_c$ fit. Bottom: distribution of the fitted diffusion coefficient in the case of $E_c=1$\,PeV and $E_c$ left free to vary, for a combined fit to H.E.S.S. and HAWC data. Solid lines indicate the value ranges for the B/C diffusion coefficient and the statistical uncertainty of HAWC diffusion coefficient obtained in \cite{HAWC17}.}
\label{fig:fitResults}
\end{figure}

The fitted diffusion coefficients are comparable with the one obtained by HAWC \citep{HAWC17} as shown in Fig. \ref{fig:fitResults}. The diffusion coefficients derived from the Boron-to-Carbon ratio (labelled B/C diffusion hereafter) are obtained under the assumption that the galactic magnetic halo size is $1$\,kpc, $4$\,kpc, and $16$\,kpc respectively for the three values indicated \citep{genolini:hal-02101560}. The H.E.S.S. result confirms a significantly lower diffusion coefficient than that obtained with the cosmic ray Boron-to-Carbon ratio. Moreover, the mean statistical uncertainty of the fits improved from $27 \%$ with HAWC only to $17 \%$ with the combination of H.E.S.S. and HAWC. 

A comparison is performed between models both with an energy cut-off fixed to $1$\,PeV and when leaving the energy cut-off free to vary. This value is chosen as the minimal cut-off value that does not significantly modify the spectral shape in the energy ranges of H.E.S.S. and HAWC. The cut-off is defined as significant if the model fit leaving $E_c$ free is able to provide a good fit (p-value $> 0.003$) and if the chi-square difference with the $E_c$ fixed model is found to be higher than $25$ (which corresponds to $5$ standard deviations). 

For the H.E.S.S. and HAWC combined data, $53$ models passed this criterion showing preference for a sub-PeV cut-off. Fig. \ref{fig:fitResults} shows the fitted energy cut-off compared to the power law index of the injection spectrum and the fitted diffusion coefficient for good fits obtained with an energy cut-off either fixed to 1\,PeV or left free to vary. A correlation can be observed between the spectral index and the energy cut-off in the top panel of Fig. \ref{fig:fitResults}, with a Pearson correlation coefficient of $0.64$. A hard injection spectrum, with index lower than $2$, implies the presence of a sub-PeV energy cut-off to reproduce the combined H.E.S.S. and HAWC dataset. 

\begin{figure}
\centering 
\includegraphics[width=\columnwidth]{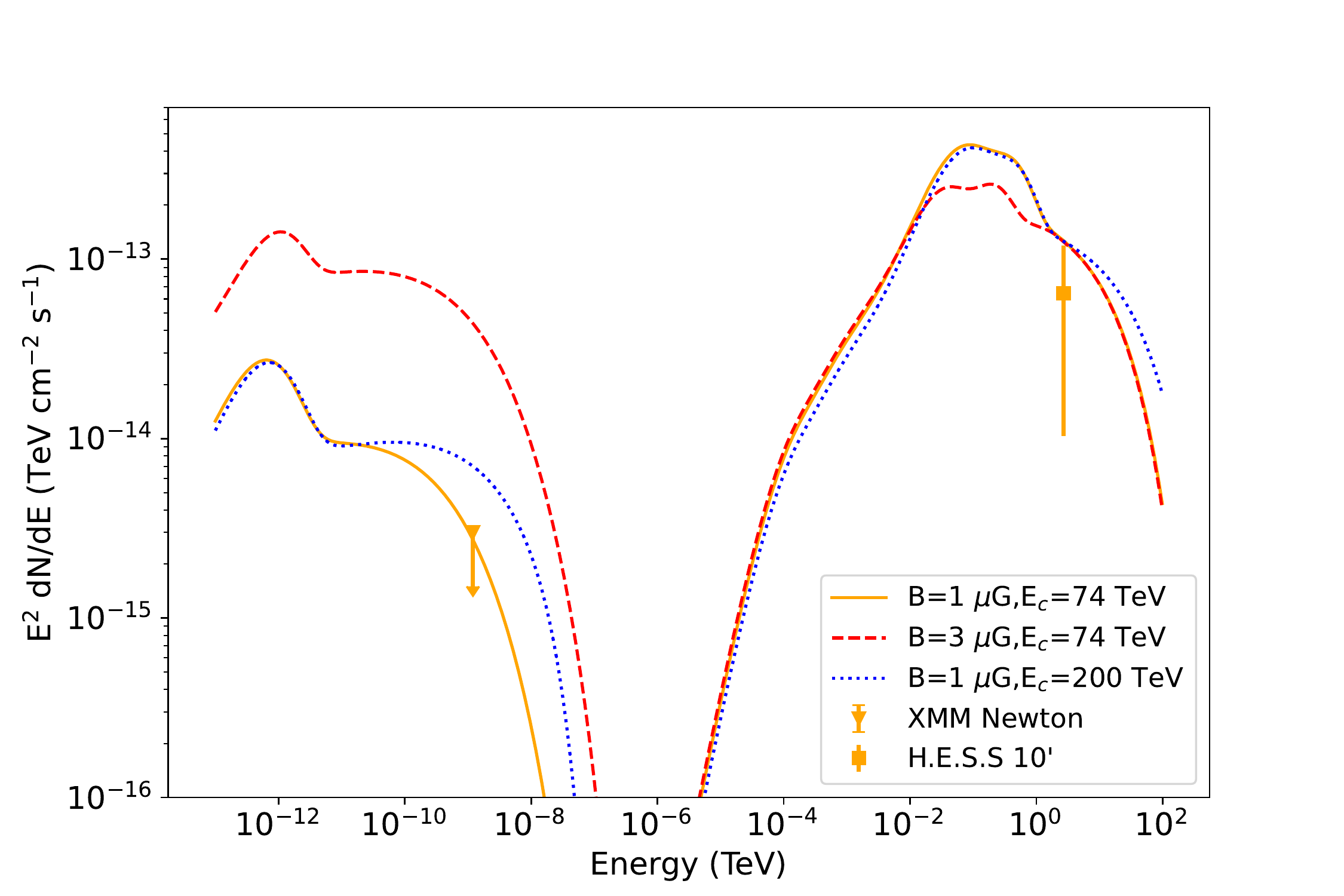}
\caption{Comparison of the model SED with the XMM-Newton upper limit for different parameters. A flux point for H.E.S.S. corresponding to the innermost radii $\lesssim10'$ is shown for comparison. }
\label{fig:fitResults_XMM}
\end{figure}

The expected synchrotron flux in the $10'$ region around the pulsar from the good fit models is compared with the XMM-Newton upper limits with both a free cut-off and a fixed 1\,PeV cut-off. 
For the models defined as a good fit of the H.E.S.S. and HAWC data, those with a fixed $E_c$ of 1\,PeV result in a synchrotron flux exceeding the upper limit. This leads  to their exclusion, while for the free cut-off, the models with a magnetic field of $1$\,$\mu$G and a high-energy cut-off lower than $75$\,TeV are able to fulfil the XMM-Newton constraint. 
The impact of these two parameters on the SED is presented in the Fig. \ref{fig:fitResults_XMM}. A value higher than $1$\,$\mu$G overshoots the X-ray upper limits, as well as a higher injection cut-off energy than $75$\,TeV. A flux point for H.E.S.S. derived from the innermost radial bin of Fig. \ref{fig:profile}, a comparable angular scale of $\sim10'$ is also shown for comparison. We note that due to the background normalisation applied, the absolute flux value is likely underestimated. 

\begin{figure}
\centering 
\includegraphics[width=\columnwidth]{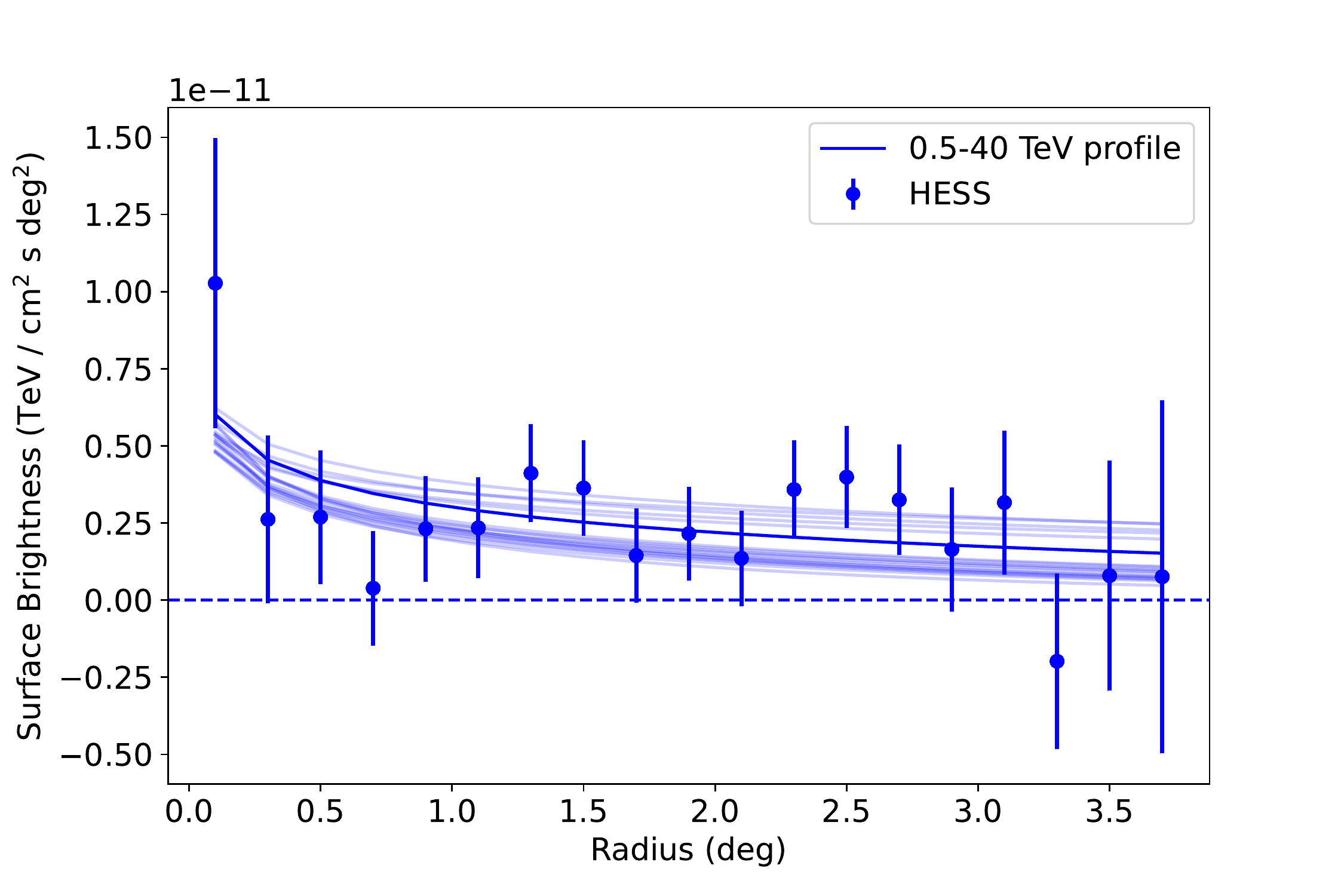}

\includegraphics[width=\columnwidth]{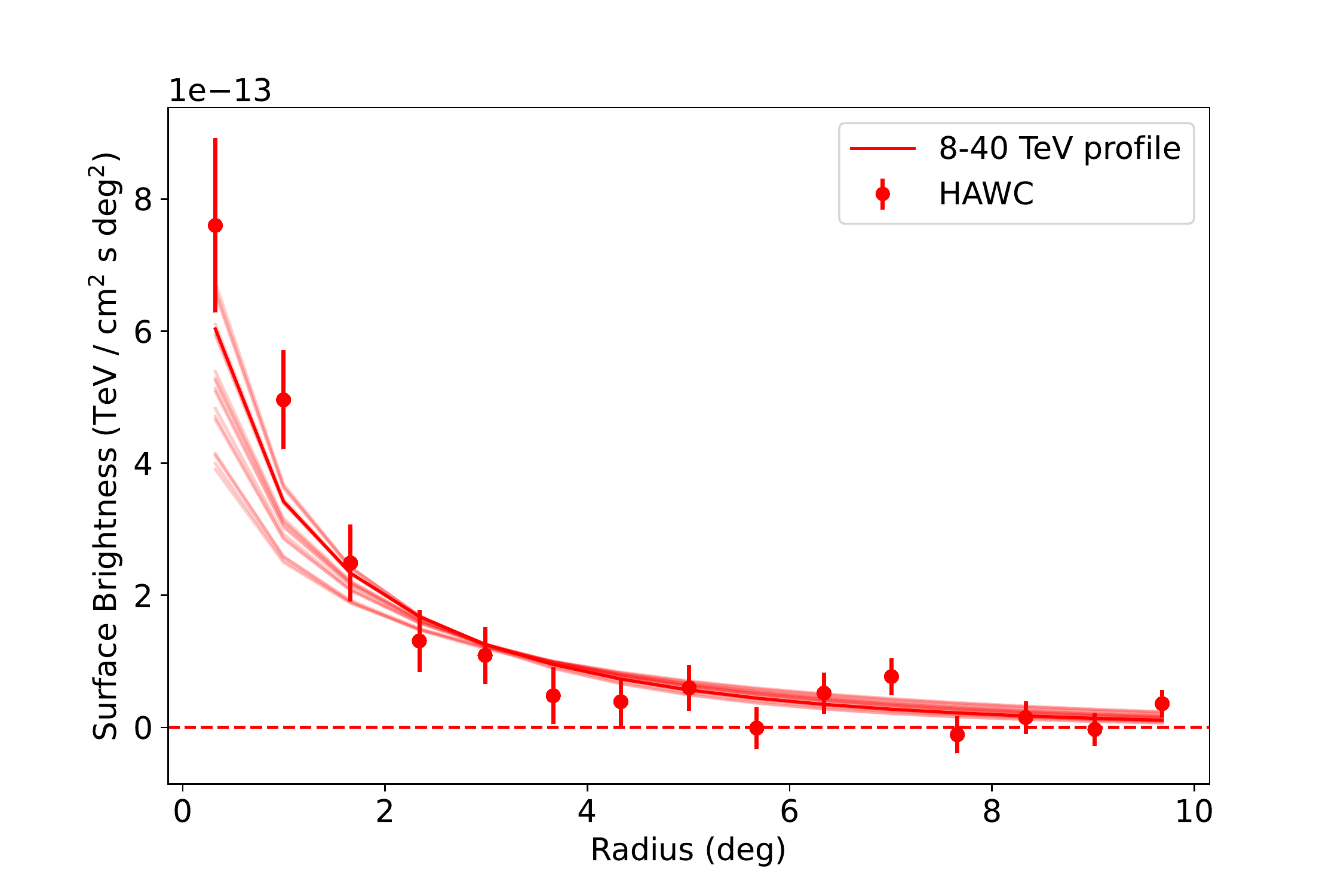}

\includegraphics[width=\columnwidth]{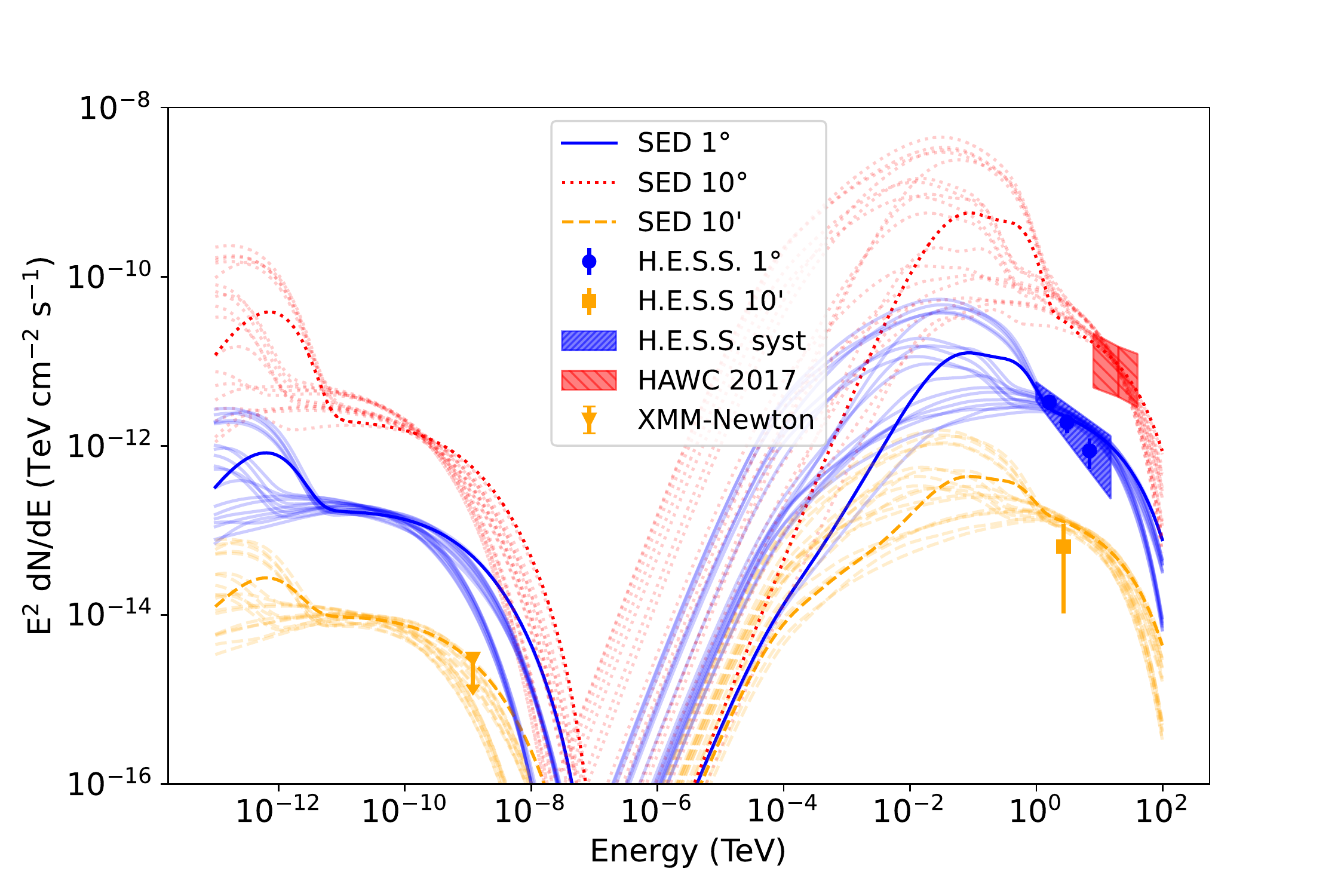}

\caption{Comparison of the fitted SED and profiles with data for the models with p-value $> 0.003$. The best fit is obtained for $n = 4.5$, $\eta = 0.1$, $\alpha = 1.8$, $\delta = 1.0$, $B = 1\mu$G and is highlighted compared to other models, with fitted parameters of $D_0 = 7.6^{+1.5}_{-1.2} \times 10^{27}$ cm$^2$s$^{-1}$ and $E_c = 74^{+17}_{-11}$ TeV. The uncertainty bands for H.E.S.S. indicates the systematic uncertainty with statistical uncertainties indicated by the error bars, while for HAWC the uncertainty band represents the systematic and statistical uncertainty. }

\label{fig:fitComparison}
\end{figure}

As an example, the best fit from all the scanned parameters with HAWC and H.E.S.S. data in terms of p-value is shown in the Fig. \ref{fig:fitComparison} compared with data. The list of scanned models that fulfiled the good fit and XMM-Newton constraints is given in the appendix \ref{sec:Results_Scan}.

\section{Discussion}
\label{sec:discuss}

This H.E.S.S. detection of extended TeV emission around the Geminga pulsar confirms the results reported by the water Cherenkov detectors Milagro, HAWC, LHAASO, and Tibet-AS$\gamma$ with IACTs for the first time. 
Demonstrating this detection with IACT analysis techniques, potentially opens up the source class of highly extended `halos' of TeV $\gamma$-ray emission around middle-aged pulsars to detailed studies by IACTs \citep{Linden2017PhRvD..96j3016L,Giacinti2020}. In contrast to the known population of TeV pulsar wind nebulae, for which the TeV structures are already known to be typically larger than the X-ray synchrotron components \citep{1997MNRAS.291..162A}, in pulsar halos, it is thought that the emission originates from escaped particles diffusing away from the canonical PWN. 

Geminga is unambiguously in the halo stage of PWN environmental evolution and as such is the first clear example of extended halo-like TeV emission to be detected with H.E.S.S. or by IACT-based facilities in general \citep{Giacinti2020}. Previously, the limited field of view, pointed observations, and restrictive background estimation approaches of IACTs have prohibited such a measurement. However, this paper demonstrates that these aspects can be partly overcome, although a full characterisation of the $\gamma$-ray emission around Geminga remains challenging. 
Given the systematic uncertainties, no significant asymmetry in the morphology of the $\gamma$-ray emission is found. 
If an asymmetry were to be confirmed at a significant level, it could provide valuable information and constraints on the level of magnetic turbulence in the region \citep{Turbulence2018MNRAS.479.4526L}.

Given the observed extent of $\gamma$-ray emission around Geminga by the survey instruments HAWC and Fermi-LAT of 20-30\,pc and $\sim100$\,pc respectively, we can anticipate that the true extent of TeV $\gamma$-ray emission within the H.E.S.S. energy range is $\sim50$\,pc, or roughly double the current H.E.S.S. field of view, following the estimate shown in Fig. \ref{fig:model_profile}.
In contrast, the observed extent in other wavebands is just $~2'$ in X-ray, $10''$ in radio or a mere few arcseconds in optical \citep{deLuca2006A&A...445L...9D,Radio2011MNRAS.416L..45P,SubaruOptical2006A&A...448..313S}. It is hence apparent that the emission detected in these wavelength bands are produced by different regions within the PWN-halo system. 
Nevertheless, given the $\gamma$-ray evidence for a halo of escaped electrons far larger than the region indicated for energetic electrons in X-ray and radio observations to date, one may expect that a larger region of much weaker low surface brightness synchrotron emission exists, originating from this electron population. 
Faint X-ray emission associated with the wider halo of escaped electrons may be identifiable with instruments sensitive to low surface brightness emission, such as eROSITA \citep{erosita2021A&A...647A...1P}.

With the revised energy loss description and analytical solution from \cite{diMauroFermi2019PhRvD.100l3015D} as described in section \ref{sec:model}, a value for $D_0$ compatible with that of the HAWC analysis \citep{HAWC17} is obtained. Three turbulence regimes are probed, Kolmogorov and Kraichnan turbulence, as well as Bohm diffusion, employing the diffusion index $\delta \in [0.3, 0.6, 1.]$ respectively \citep{Kolmogorov10.2307/51980,Kraichnan,BohmDiff}. No preferred value is found with the joint fit of H.E.S.S. and HAWC data, as well as no preference between the ambient magnetic field, injection index, and braking index over the scanned values. We find that the efficiency with which energy is converted to electrons has to be higher than $1 \%$ giving a direct lower limit to the proportion of electrons that escape into the pulsar wind nebula over the history of the pulsar. Even if the spectral index is not directly constrained, a spectral index lower than $2$ implies the presence of a sub-PeV energy cut-off in order to explain the combined H.E.S.S. and HAWC data. 

Considerable systematic uncertainties are also inherent in this analysis, as demonstrated by the different background estimation methods used. It was not possible to measure the true extent of the $\gamma$-ray emission around the Geminga pulsar, from the currently available H.E.S.S. data.  Caution must therefore be urged in the interpretation of the diffusion model fit in particular: although the fit is performed simultaneously to both the radial profile and spectrum, the radial extent of detected emission is limited by the detector field of view. Given the uncertainty in where the level of background emission is reached (and necessary assumptions made for background normalisation), this does not enable us to make an absolute measurement of the flux or an assumption-free measurement of the diffusion coefficient. 

Our study of the X-ray upper limits in a $10'$ region around the pulsar allows us to conclude that the magnetic field, in a one diffusion zone scenario, and assuming a constant magnetic field over the X-ray to $\gamma$-ray scale, has to be lower than $1\,\mu G$ in the absence of a sub-PeV energy cut-off. 
An energy cut-off lower than $75$\,TeV is needed to account for a magnetic field of $1\,\mu G$; the absence of such a cut-off would imply a sub 1\,$\mu G$ magnetic field leading to a tension with the typical value for the normalisation of the magnetic field of the interstellar medium ($2.3\,\mu{\rm G} \lesssim B \lesssim 6.1\,\mu {\rm G}$ \citealt{Delahaye10}). 
This can be seen as another argument favouring the presence of a sub-PeV cut-off in the pulsar injection. The scenarios where the magnetic field is $\sim 1\,\mu G$ and a strong cut-off is present cannot be disentangled from a sub-$\mu G$ magnetic field, but an extension towards higher energies of the current $\gamma$-ray observations would be able to distinguish these two scenarios. 

Several models have considered diffusion as the dominant transport mechanism for trapped particles within PWNe \citep{2012ApJ...752...83T,2016MNRAS.460.4135P}. However in older, more evolved systems the halo of energetic particles probes properties of the ambient medium. Initially, it was surprising that diffusion at larger distances away from the pulsar continued to be considerably below the Galactic average value. If this value is representative of properties of the intervening ISM, it would question the paradigm of pulsars as the origin of the cosmic-ray positron excess. Several theories have been proposed to reconcile these aspects, the most popular thus far being that the region of slow diffusion is constrained to the region around the pulsar in which accelerated electrons continue to have significant influence on the ISM \citep{Evoli18,Fang2018ApJ...863...30F,FangSlowDiff2019MNRAS.488.4074F,Profumo2018PhRvD..97l3008P}. At larger distances, the lower particle energy densities reduce the influence, such that the halo emission gradually decreases to join smoothly with the ISM. 
Alternatively, it has been noted that although the implied diffusion coefficient is lower than the Galactic average value, it remains consistent with predictions of a spiral arm model with non-uniform diffusion coefficient throughout the Galaxy \citep{Tang2019MNRAS.484.3491T}.

In the future, observing these types of large extended sources will become more feasible with the Cherenkov Telescope Array (CTA, the next-generation IACT $\gamma$-ray observatory) due to the anticipated larger telescope field of view, of up to $\sim 8-9^\circ$ towards the highest energies \citep{Acharya13}. 
Additionally, divergent pointing strategies are foreseen to cover a larger region of the sky at a single observation instant, using the same number of telescopes but trading an increased field of view for reduced sensitivity \citep{2021Univ....7..421DeNaurois,Donini:2019Kx}. Whilst this is envisaged in particular for observations of transient phenomena where the location may be poorly constrained, there are obvious advantages of employing this mode to observe extended `halo' phenomena. 
With these approaches and a much larger number of available telescopes, several of the challenges limiting the capabilities of current IACT arrays such as H.E.S.S. to observe pulsar halos and comparable phenomena at TeV energies will be overcome with CTA. Further survey instruments are also foreseen, that may increase the number of confirmed pulsar halos dramatically in the near future \citep{2022NatAs...6..199L}.

\section{Conclusions}

With this study, H.E.S.S. confirms the presence of extended very-high-energy $\gamma$-ray emission around the Geminga pulsar, on angular scales reaching at least $3.2^\circ$ radius. Two different methods of background estimation were used to evaluate the systematic uncertainties of the measurement; the field-of-view background method and the On-Off background method, with two independent lists of Off runs for the latter. No evidence for statistically significant asymmetries to the emission or energy-dependent morphology is found. Within a $1^\circ$ radius region around the pulsar, a spectral analysis obtained a flux normalisation at 1\,TeV of $(2.8\pm0.7)\times10^{-12}$\,cm$^{-2}$s$^{-1}$TeV$^{-1}$ with a best-fit power law spectral index of $2.76\pm 0.22$. We fitted an electron diffusion model jointly to the H.E.S.S. data combined with HAWC and compare to XMM-Newton, taking the different spectral extraction regions into account. Thanks to the few-arcmin angular resolution of H.E.S.S., it was possible to extract a flux point for the innermost 10' radius region around the pulsar, corresponding to the XMM-Newton upper limit. Due to the large number of free variables in the fit, a parameter scan was conducted to cycle over specific values for several quantities, whilst leaving the diffusion coefficient normalisation free. For the best-fit model, a normalisation of the diffusion coefficient of $D_0 = 7.6^{+1.5}_{-1.2} \times 10^{27}$ cm$^2$s$^{-1}$ is preferred at an electron energy of 100\,TeV, as well as a cut off energy of the electron spectrum $E_c=74^{+17}_{-11}$\,TeV. This value is considerably lower than the Galactic average, yet consistent with results obtained by the HAWC collaboration \citep{HAWC17}.  
We find that the mean statistical uncertainty on the diffusion coefficient obtained from our model fit is 17\% for the joint fit of H.E.S.S. and HAWC data, whereas it is 27\% for the fit to HAWC data only. 
 
This is a challenging analysis: due to the $\gamma$-ray emission extending beyond the available sky region, limitations apply such as a lower limit rather than absolute measurement of the size, and a relative rather than absolute flux measurement, likely underestimating the true $\gamma$-ray flux.  Despite these caveats, we anticipate that this detection and study of extended $\gamma$-ray emission around the Geminga pulsar paves the way for further detailed studies of pulsar halos with current and future generation IACT facilities. 

\section*{Acknowledgements}
\label{sec:acknowledge}
\textit{
The support of the Namibian authorities and of the University of
Namibia in facilitating the construction and operation of H.E.S.S.
is gratefully acknowledged, as is the support by the German
Ministry for Education and Research (BMBF), the Max Planck Society,
the German Research Foundation (DFG), the Helmholtz Association,
the Alexander von Humboldt Foundation, the French Ministry of
Higher Education, Research and Innovation, the Centre National de
la Recherche Scientifique (CNRS/IN2P3 and CNRS/INSU), the
Commissariat à l’énergie atomique et aux énergies alternatives
(CEA), the U.K. Science and Technology Facilities Council (STFC),
the Irish Research Council (IRC) and the Science Foundation Ireland
(SFI), the Knut and Alice Wallenberg Foundation, the Polish
Ministry of Education and Science, agreement no. 2021/WK/06, the
South African Department of Science and Technology and National
Research Foundation, the University of Namibia, the National
Commission on Research, Science \& Technology of Namibia (NCRST),
the Austrian Federal Ministry of Education, Science and Research
and the Austrian Science Fund (FWF), the Australian Research
Council (ARC), the Japan Society for the Promotion of Science, the
University of Amsterdam and the Science Committee of Armenia grant
21AG-1C085. We appreciate the excellent work of the technical
support staff in Berlin, Zeuthen, Heidelberg, Palaiseau, Paris,
Saclay, T\"ubingen and in Namibia in the construction and operation
of the equipment. This work benefited from services provided by the
H.E.S.S. Virtual Organisation, supported by the national resource
providers of the EGI Federation.
}

\bibliographystyle{aa-package/bibtex/aa}
\bibliography{Gemingarefs}

\begin{appendix}
\label{sec:appendix}

\section{Diffusion model} 
\label{sec:modelextra}
\subsection{Energy losses}
\label{sec:EnergyLosses}

The energy losses are assumed to be dominated by synchrotron losses and Inverse Compton losses. For the latter, we use the model developed in \cite{Delahaye10} (equation 32), which approximates the intermediate regime between Thomson and Klein-Nishina regimes. 
For the Inverse Compton losses, the three thermal radiation fields considered have the following properties \citep{1998ApJ...493..694M}: Cosmic Microwave Background 
($T = 2.7$K, $U_{rad} = 0.26$ eV\,cm$^{-3}$), Infra-Red ($T = 20$K, $U_{rad} = 0.3$ eV\,cm$^{-3}$) and Optical ($T = 5000$K, $U_{rad} = 0.3$ eV\,cm$^{-3}$).

The total energy losses are computed as: 
\begin{linenomath*}
\begin{equation}
b(E) = b_{\mathrm{sync}}(E) + b_{\mathrm{CMB}}(E) + b_{\mathrm{IR}}(E) + b_{\mathrm{Opt}}(E)~,
\end{equation}
\end{linenomath*}
including losses due to synchrotron $b_{\rm sync}$ and IC scattering on the three aforementioned radiation fields. Due to the different temperatures and energy densities, the transition between the Thomson and Klein-Nishina IC scattering regimes occurs at different energies for the three radiation fields. 

\begin{figure}
    \centering
    \includegraphics[width=\columnwidth]{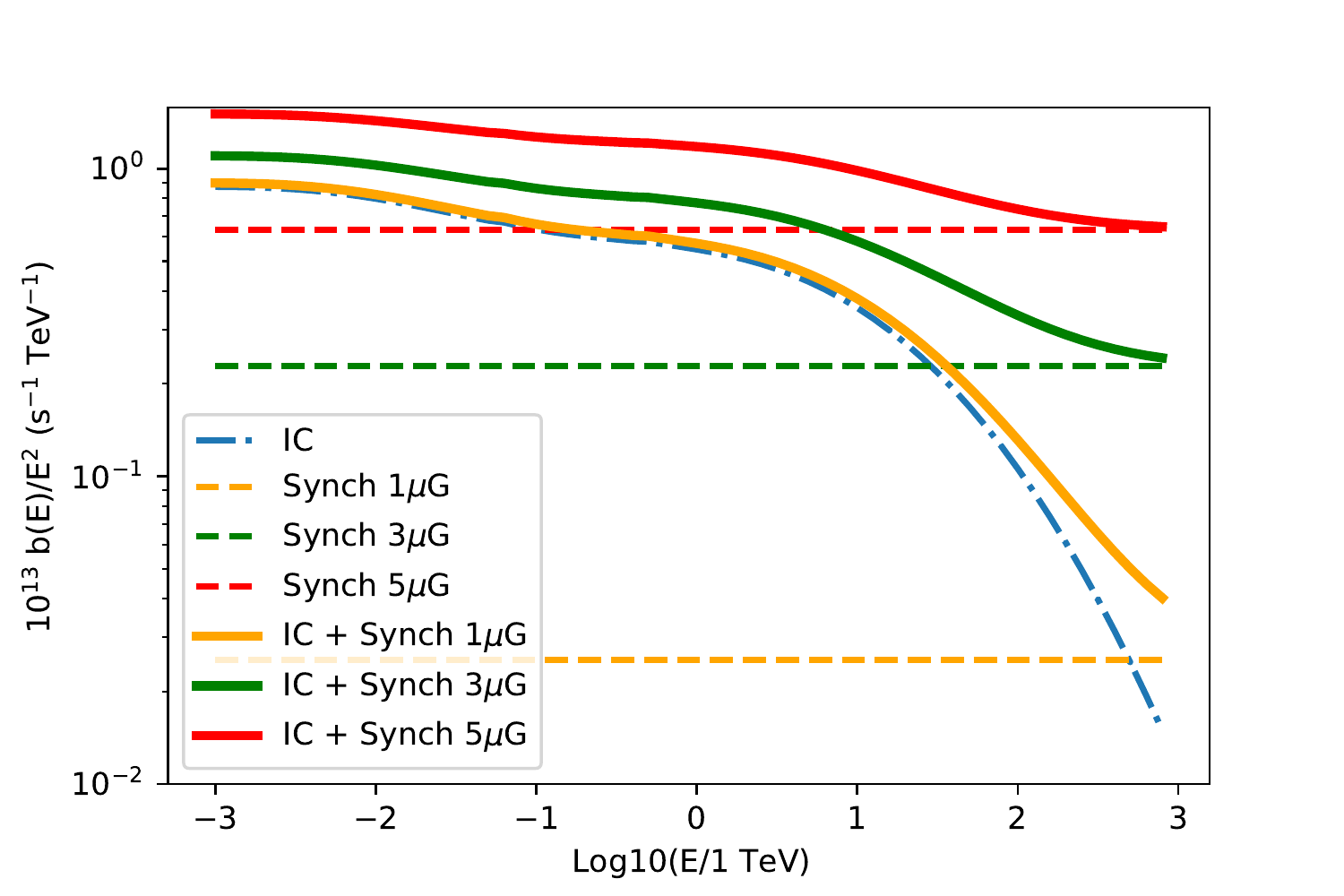}
    \caption{Energy losses from IC and synchrotron processes for different magnetic fields used in the diffusion model. IC is the total inverse Compton losses over all three radiation fields considered.} 
    \label{fig:rad_fields}
\end{figure}

With increasing magnetic field strength, the energy range for which Synchrotron losses dominate over inverse Compton losses also increases, as shown in Fig. \ref{fig:rad_fields}.

\subsection{Pulsar properties}
\label{appendix:pulsar}
The pulsar luminosity as a function of time is defined as:
\begin{linenomath*}
\begin{equation}
    L(t) = L_0 (1 + t/\tau_0)^{-(n+1)/(n-1)}~,
    \label{eq:lum}
\end{equation}
\end{linenomath*}
where $L_0$ is the initial luminosity of the pulsar, $\tau_0$ the initial spin-down timescale, and $n$ is the braking index. The pulsar period as a function of time is: 
\begin{linenomath*}
\begin{equation}
    P(t) = P_0 (1 + t/\tau_0)^{1/(n-1)}~,
    \label{eq:pt}
\end{equation}
\end{linenomath*}
and the pulsar age $T_*$ is expressed as: 
\begin{linenomath*}
\begin{equation}
    T_*= T_c \frac{2}{n-1} (1-P_0/P_*)^{n-1}~,
    \label{eq:age}
\end{equation}
\end{linenomath*}
with $T_c$ the characteristic age and $P_*$ the actual pulsar period. It is important to note here that $T_*\simeq T_c$ for $n=3$ and $P_* \gg P_0$ if the pulsar is old enough to have had a significant increase of its rotation period. Given these previous equations, we can retrieve $\tau_0$ solving the equations \eqref{eq:pt} and \eqref{eq:age} with $P(T_*) = P_*$ :
\begin{linenomath*}
\begin{equation}
    \tau_0 = \frac{T_*}{(P_*/P_0)^{n-1}-1}~.
    \label{eq:tau0}
\end{equation}
\end{linenomath*}

\newpage

\section{Results for the SED and profile fit}
\label{sec:Results_Scan}

\begin{table}[h!]
    \caption{Fit results on H.E.S.S. and HAWC SED and profiles of the scanned parameter combinations with a p-value better than 3 standard deviations and fulfiling the XMM-Newton constraints. }
    \label{tab:Results_Scan}
    \renewcommand{\arraystretch}{1.3}
    \centering
    \begin{tabular}[width=\columnwidth]{cccccccc}
    $n$ & $\eta$ & $\alpha$ & $E_c$ & $\delta$ & $D_0$  & p-value & Chi2/ndof \\
    & & & TeV &  & $10^{27} {\rm cm}^2 {\rm s}^{-1}$ & & \\
    \hline
    $4.5$ & $0.1$ & $1.8$ & $74^{+17}_{-11}$ & $1.0$ & $7.6^{+1.5}_{-1.2}$ & $0.37$ & $1.07$ \\
    $4.5$ & $0.5$ & $2.0$ & $47^{+6}_{-5}$ & $0.3$ & $5.7^{+1.1}_{-0.9}$ & $0.35$ & $1.07$ \\
    $4.5$ & $0.5$ & $2.0$ & $53^{+9}_{-6}$ & $0.6$ & $10.0^{+1.9}_{-1.6}$ & $0.19$ & $1.21$ \\
    $3.0$ & $0.5$ & $2.0$ & $45^{+6}_{-5}$ & $0.3$ & $4.6^{+0.9}_{-0.8}$ & $0.17$ & $1.24$ \\
    $3.0$ & $0.5$ & $2.0$ & $49^{+7}_{-5}$ & $0.6$ & $7.3^{+1.4}_{-1.2}$ & $0.15$ & $1.26$ \\ 
    $1.5$ & $0.5$ & $2.0$ & $45^{+6}_{-5}$ & $0.3$ & $4.5^{+0.9}_{-0.8}$ & $0.14$ & $1.28$\\
    $1.5$ & $0.5$ & $2.0$ & $49^{+7}_{-5}$ & $0.6$ & $7.1^{+1.3}_{-1.1}$ & $0.13$ & $1.28$ \\
    $3.0$ & $0.1$ & $1.8$ & $71^{+15}_{-10}$ & $1.0$ & $6.5^{+1.2}_{-1.0}$ & $0.13$ & $1.29$ \\
    $4.5$ & $0.5$ & $1.8$ & $27^{+3}_{-2}$ & $0.3$ & $12.1^{+2.0}_{-1.7}$ & $0.11$ & $1.32$ \\
    $3.0$ & $0.5$ & $1.8$ & $25^{+2}_{-2}$ & $0.3$ & $9.1^{+1.7}_{-1.4}$ & $0.08$ & $1.36$ \\
    $1.5$ & $0.5$ & $2.0$ & $57^{+10}_{-7}$ & $1.0$ & $14.0^{+2.5}_{-2.1}$ & $0.08$ & $1.37$ \\
    $3.0$ & $0.5$ & $2.0$ & $57^{+10}_{-7}$ &  $1.0$ & $14.4^{+2.6}_{-2.2}$ & $0.08$ & $1.37$ \\
    $1.5$ & $0.5$ & $1.8$ & $26^{+2}_{-2}$ & $0.3$ & $8.9^{+1.7}_{-1.4}$ & $0.07$ & $1.39$\\
    $3.0$ & $0.5$ & $1.8$ & $27^{+3}_{-2}$ & $0.6$ & $15.6^{+2.8}_{-2.3}$ & $0.05$ & $1.43$ \\
    $1.5$ & $0.5$ & $1.8$ & $27^{+3}_{-2}$ & $0.6$ & $15.1^{+2.7}_{-2.3}$ & $0.05$ & $1.45$ \\
    $4.5$ & $0.5$ & $2.0$ & $62^{+14}_{-10}$ & $1.0$ & $20.7^{+4.5}_{-3.7}$ & $0.04$ & $1.47$ \\
    $4.5$ & $0.5$ & $1.8$ & $30^{+3}_{-3}$ & $0.6$ & $23.9^{+3.9}_{-3.3}$ & $0.03$ & $1.52$ \\
    $1.5$ & $0.5$ & $1.8$ & $31^{+4}_{-3}$ & $1.0$ & $32.0^{+5.4}_{-4.6}$ & $0.02$ & $1.59$ \\
    $3.0$ & $0.5$ & $1.8$ & $31^{+4}_{-3}$ & $1.0$ & $33.5^{+5.7}_{-4.9}$ & $0.02$ & $1.59$ \\
    \end{tabular}
    \tablefoot{The magnetic field is set to $B=1\,\mu$G for all combinations listed.}
\end{table}

\end{appendix}
\end{document}

%% file: authors_Detection_of_GemingaHESS_AA.tex
\author{H.E.S.S. Collaboration
\and F.~Aharonian \inst{\ref{DIAS},\ref{MPIK}}
\and F.~Ait~Benkhali \inst{\ref{LSW}}
\and J.~Aschersleben \inst{\ref{Groningen}}
\and H.~Ashkar \inst{\ref{LLR}}
\and M.~Backes \inst{\ref{UNAM},\ref{NWU}}
\and V.~Barbosa~Martins \inst{\ref{DESY}}
\and R.~Batzofin \inst{\ref{Wits}}
\and Y.~Becherini \inst{\ref{APC},\ref{Linnaeus}}
\and D.~Berge \inst{\ref{DESY},\ref{HUB}}
\and K.~Bernl\"ohr \inst{\ref{MPIK}}
\and B.~Bi \inst{\ref{IAAT}}
\and M.~B\"ottcher \inst{\ref{NWU}}
\and C.~Boisson \inst{\ref{LUTH}}
\and J.~Bolmont \inst{\ref{LPNHE}}
\and J.~Borowska \inst{\ref{HUB}}
\and M.~Bouyahiaoui \inst{\ref{MPIK}}
\and F.~Bradascio \inst{\ref{CEA}}
\and R.~Brose \inst{\ref{DIAS}}
\and F.~Brun \inst{\ref{CEA}}
\and B.~Bruno \inst{\ref{ECAP}}
\and T.~Bulik \inst{\ref{UWarsaw}}
\and C.~Burger-Scheidlin \inst{\ref{DIAS}}
\and F.~Cangemi \inst{\ref{LPNHE}}
\and S.~Caroff \inst{\ref{LAPP}}$^{,}$\thanks{Corresponding authors;\newline\email{\href{mailto:contact.hess@hess-experiment.eu}{contact.hess@hess-experiment.eu}}}
\and S.~Casanova \inst{\ref{IFJPAN}}
\and J.~Celic \inst{\ref{ECAP}}
\and M.~Cerruti \inst{\ref{APC}}
\and P.~Chambery \inst{\ref{CENBG}}
\and T.~Chand \inst{\ref{NWU}}
\and S.~Chandra \inst{\ref{NWU}}
\and A.~Chen \inst{\ref{Wits}}
\and J.~Chibueze \inst{\ref{NWU}}
\and O.~Chibueze \inst{\ref{NWU}}
\and G.~Cotter \inst{\ref{Oxford}}
\and J.~Damascene~Mbarubucyeye \inst{\ref{DESY}}
\and J.~Devin \inst{\ref{LUPM}}
\and A.~Djannati-Ata\"i \inst{\ref{APC}}
\and A.~Dmytriiev \inst{\ref{NWU}}
\and K.~Egberts \inst{\ref{UP}}
\and S.~Einecke \inst{\ref{Adelaide}}
\and J.-P.~Ernenwein \inst{\ref{CPPM}}
\and K.~Feijen \inst{\ref{Adelaide}}
\and G.~Fichet~de~Clairfontaine \inst{\ref{LUTH}}
\and M.~Filipovic \inst{\ref{Sydney}}
\and G.~Fontaine \inst{\ref{LLR}}
\and M.~F\"u{\ss}ling \inst{\ref{DESY}}
\and S.~Funk \inst{\ref{ECAP}}
\and S.~Gabici \inst{\ref{APC}}
\and Y.A.~Gallant \inst{\ref{LUPM}}
\and S.~Ghafourizadeh \inst{\ref{LSW}}
\and G.~Giavitto \inst{\ref{DESY}}
\and L.~Giunti \inst{\ref{APC},\ref{CEA}}
\and D.~Glawion \inst{\ref{ECAP}}
\and J.F.~Glicenstein \inst{\ref{CEA}}
\and P.~Goswami \inst{\ref{NWU}}
\and G.~Grolleron \inst{\ref{LPNHE}}
\and M.-H.~Grondin \inst{\ref{CENBG}}
\and L.~Haerer \inst{\ref{MPIK}}
\and M.~Haupt \inst{\ref{DESY}}
\and G.~Hermann \inst{\ref{MPIK}}
\and J.A.~Hinton \inst{\ref{MPIK}}
\and W.~Hofmann \inst{\ref{MPIK}}
\and T.~L.~Holch \inst{\ref{DESY}}
\and M.~Holler \inst{\ref{Innsbruck}}
\and D.~Horns \inst{\ref{UHH}}
\and Zhiqiu~Huang \inst{\ref{MPIK}}
\and M.~Jamrozy \inst{\ref{UJK}}
\and F.~Jankowsky \inst{\ref{LSW}}
\and V.~Joshi \inst{\ref{ECAP}}
\and I.~Jung-Richardt \inst{\ref{ECAP}}
\and E.~Kasai \inst{\ref{UNAM}}
\and K.~Katarzy{\'n}ski \inst{\ref{NCUT}}
\and B.~Kh\'elifi \inst{\ref{APC}}
\and W.~Klu\'{z}niak \inst{\ref{NCAC}}
\and Nu.~Komin \inst{\ref{Wits}}
\and K.~Kosack \inst{\ref{CEA}}
\and D.~Kostunin \inst{\ref{DESY}}
\and R.G.~Lang \inst{\ref{ECAP}}
\and S.~Le~Stum \inst{\ref{CPPM}}
\and F.~Leitl \inst{\ref{ECAP}}
\and A.~Lemi\`ere \inst{\ref{APC}}
\and M.~Lemoine-Goumard \inst{\ref{CENBG}}
\and J.-P.~Lenain \inst{\ref{LPNHE}}
\and F.~Leuschner \inst{\ref{IAAT}}
\and T.~Lohse \inst{\ref{HUB}}
\and A.~Luashvili \inst{\ref{LUTH}}
\and I.~Lypova \inst{\ref{LSW}}
\and J.~Mackey \inst{\ref{DIAS}}
\and D.~Malyshev \inst{\ref{IAAT}}
\and V.~Marandon \inst{\ref{MPIK}}
\and P.~Marchegiani \inst{\ref{Wits}}
\and A.~Marcowith \inst{\ref{LUPM}}
\and P.~Marinos \inst{\ref{Adelaide}}
\and G.~Mart\'i-Devesa \inst{\ref{Innsbruck}}
\and R.~Marx \inst{\ref{LSW}}
\and G.~Maurin \inst{\ref{LAPP}}
\and P.J.~Meintjes \inst{\ref{UFS}}
\and M.~Meyer \inst{\ref{UHH}}
\and A.~Mitchell  \inst{\ref{ECAP}}$^{,}$\footnotemark[1]
\and R.~Moderski \inst{\ref{NCAC}}
\and L.~Mohrmann \inst{\ref{MPIK}}
\and A.~Montanari \inst{\ref{CEA}}
\and E.~Moulin \inst{\ref{CEA}}
\and J.~Muller \inst{\ref{LLR}}
\and K.~Nakashima \inst{\ref{ECAP}}
\and M.~de~Naurois \inst{\ref{LLR}}
\and J.~Niemiec \inst{\ref{IFJPAN}}
\and A.~Priyana~Noel \inst{\ref{UJK}}
\and P.~O'Brien \inst{\ref{Leicester}}
\and S.~Ohm \inst{\ref{DESY}}
\and L.~Olivera-Nieto \inst{\ref{MPIK}}
\and E.~de~Ona~Wilhelmi \inst{\ref{DESY}}
\and M.~Ostrowski \inst{\ref{UJK}}
\and S.~Panny \inst{\ref{Innsbruck}}
\and M.~Panter \inst{\ref{MPIK}}
\and R.D.~Parsons \inst{\ref{HUB}}
\and G.~Peron \inst{\ref{APC}}
\and D.A.~Prokhorov \inst{\ref{Amsterdam}}
\and G.~P\"uhlhofer \inst{\ref{IAAT}}
\and A.~Quirrenbach \inst{\ref{LSW}}
\and A.~Reimer \inst{\ref{Innsbruck}}
\and O.~Reimer \inst{\ref{Innsbruck}}
\and M.~Renaud \inst{\ref{LUPM}}
\and B.~Reville \inst{\ref{MPIK}}
\and F.~Rieger \inst{\ref{MPIK}}
\and G.~Rowell \inst{\ref{Adelaide}}
\and B.~Rudak \inst{\ref{NCAC}}
\and H.~Rueda Ricarte \inst{\ref{CEA}}
\and E.~Ruiz-Velasco \inst{\ref{MPIK}}
\and V.~Sahakian \inst{\ref{Yerevan}}
\and H.~Salzmann \inst{\ref{IAAT}}
\and A.~Santangelo \inst{\ref{IAAT}}
\and M.~Sasaki \inst{\ref{ECAP}}
\and F.~Sch\"ussler \inst{\ref{CEA}}
\and H.M.~Schutte \inst{\ref{NWU}}
\and U.~Schwanke \inst{\ref{HUB}}
\and J.N.S.~Shapopi \inst{\ref{UNAM}}
\and A.~Sinha \inst{\ref{LUPM}}
\and H.~Sol \inst{\ref{LUTH}}
\and A.~Specovius \inst{\ref{ECAP}}
\and S.~Spencer \inst{\ref{ECAP}}
\and {\L.}~Stawarz \inst{\ref{UJK}}
\and S.~Steinmassl \inst{\ref{MPIK}}
\and I.~Sushch \inst{\ref{NWU}}
\and H.~Suzuki \inst{\ref{Konan}}
\and T.~Takahashi \inst{\ref{KAVLI}}
\and T.~Tanaka \inst{\ref{Konan}}
\and T.~Tavernier \inst{\ref{CEA}}
\and A.M.~Taylor \inst{\ref{DESY}}
\and R.~Terrier \inst{\ref{APC}}
\and C.~Thorpe-Morgan \inst{\ref{IAAT}}
\and M.~Tsirou \inst{\ref{MPIK}}
\and N.~Tsuji \inst{\ref{RIKKEN}}
\and M.~Vecchi \inst{\ref{Groningen}}
\and C.~Venter \inst{\ref{NWU}}
\and J.~Vink \inst{\ref{Amsterdam}}
\and S.J.~Wagner \inst{\ref{LSW}}
\and R.~White \inst{\ref{MPIK}}
\and A.~Wierzcholska \inst{\ref{IFJPAN}}
\and Yu~Wun~Wong \inst{\ref{ECAP}}
\and M.~Zacharias \inst{\ref{LSW},\ref{NWU}}
\and D.~Zargaryan \inst{\ref{DIAS}}
\and A.A.~Zdziarski \inst{\ref{NCAC}}
\and A.~Zech \inst{\ref{LUTH}}
\and S.~Zouari \inst{\ref{APC}}
\and N.~\.Zywucka \inst{\ref{NWU}}
}

\institute{
Dublin Institute for Advanced Studies, 31 Fitzwilliam Place, Dublin 2, Ireland \label{DIAS} \and
Max-Planck-Institut f\"ur Kernphysik, P.O. Box 103980, D 69029 Heidelberg, Germany \label{MPIK} \and
Landessternwarte, Universit\"at Heidelberg, K\"onigstuhl, D 69117 Heidelberg, Germany \label{LSW} \and
Kapteyn Astronomical Institute, University of Groningen, Landleven 12, 9747 AD, Groningen, The Netherlands \label{Groningen} \and
Laboratoire Leprince-Ringuet, École Polytechnique, CNRS, Institut Polytechnique de Paris, F-91128 Palaiseau, France \label{LLR} \and
University of Namibia, Department of Physics, Private Bag 13301, Windhoek 10005, Namibia \label{UNAM} \and
Centre for Space Research, North-West University, Potchefstroom 2520, South Africa \label{NWU} \and
DESY, D-15738 Zeuthen, Germany \label{DESY} \and
School of Physics, University of the Witwatersrand, 1 Jan Smuts Avenue, Braamfontein, Johannesburg, 2050 South Africa \label{Wits} \and
Université de Paris, CNRS, Astroparticule et Cosmologie, F-75013 Paris, France \label{APC} \and
Department of Physics and Electrical Engineering, Linnaeus University,  351 95 V\"axj\"o, Sweden \label{Linnaeus} \and
Institut f\"ur Physik, Humboldt-Universit\"at zu Berlin, Newtonstr. 15, D 12489 Berlin, Germany \label{HUB} \and
Institut f\"ur Astronomie und Astrophysik, Universit\"at T\"ubingen, Sand 1, D 72076 T\"ubingen, Germany \label{IAAT} \and
Laboratoire Univers et Théories, Observatoire de Paris, Université PSL, CNRS, Université de Paris, 92190 Meudon, France \label{LUTH} \and
Sorbonne Universit\'e, Universit\'e Paris Diderot, Sorbonne Paris Cit\'e, CNRS/IN2P3, Laboratoire de Physique Nucl\'eaire et de Hautes Energies, LPNHE, 4 Place Jussieu, F-75252 Paris, France \label{LPNHE} \and
IRFU, CEA, Universit\'e Paris-Saclay, F-91191 Gif-sur-Yvette, France \label{CEA} \and
Friedrich-Alexander-Universit\"at Erlangen-N\"urnberg, Erlangen Centre for Astroparticle Physics, Erwin-Rommel-Str. 1, D 91058 Erlangen, Germany \label{ECAP} \and
Astronomical Observatory, The University of Warsaw, Al. Ujazdowskie 4, 00-478 Warsaw, Poland \label{UWarsaw} \and
Université Savoie Mont Blanc, CNRS, Laboratoire d'Annecy de Physique des Particules - IN2P3, 74000 Annecy, France \label{LAPP} \and
Instytut Fizyki J\c{a}drowej PAN, ul. Radzikowskiego 152, 31-342 Krak{\'o}w, Poland \label{IFJPAN} \and
Universit\'e Bordeaux, CNRS, LP2I Bordeaux, UMR 5797, F-33170 Gradignan, France \label{CENBG} \and
University of Oxford, Department of Physics, Denys Wilkinson Building, Keble Road, Oxford OX1 3RH, UK \label{Oxford} \and
Laboratoire Univers et Particules de Montpellier, Universit\'e Montpellier, CNRS/IN2P3,  CC 72, Place Eug\`ene Bataillon, F-34095 Montpellier Cedex 5, France \label{LUPM} \and
Institut f\"ur Physik und Astronomie, Universit\"at Potsdam,  Karl-Liebknecht-Strasse 24/25, D 14476 Potsdam, Germany \label{UP} \and
School of Physical Sciences, University of Adelaide, Adelaide 5005, Australia \label{Adelaide} \and
Aix Marseille Universit\'e, CNRS/IN2P3, CPPM, Marseille, France \label{CPPM} \and
School of Science, Western Sydney University, Locked Bag 1797, Penrith South DC, NSW 2751, Australia \label{Sydney} \and
Institut f\"ur Astro- und Teilchenphysik, Leopold-Franzens-Universit\"at Innsbruck, A-6020 Innsbruck, Austria \label{Innsbruck} \and
Universit\"at Hamburg, Institut f\"ur Experimentalphysik, Luruper Chaussee 149, D 22761 Hamburg, Germany \label{UHH} \and
Obserwatorium Astronomiczne, Uniwersytet Jagiello{\'n}ski, ul. Orla 171, 30-244 Krak{\'o}w, Poland \label{UJK} \and
Institute of Astronomy, Faculty of Physics, Astronomy and Informatics, Nicolaus Copernicus University,  Grudziadzka 5, 87-100 Torun, Poland \label{NCUT} \and
Nicolaus Copernicus Astronomical Center, Polish Academy of Sciences, ul. Bartycka 18, 00-716 Warsaw, Poland \label{NCAC} \and
GRAPPA, Anton Pannekoek Institute for Astronomy, University of Amsterdam,  Science Park 904, 1098 XH Amsterdam, The Netherlands \label{Amsterdam} \and
Department of Physics, University of the Free State,  PO Box 339, Bloemfontein 9300, South Africa \label{UFS} \and
Department of Physics and Astronomy, The University of Leicester, University Road, Leicester, LE1 7RH, United Kingdom \label{Leicester} \and
Yerevan Physics Institute, 2 Alikhanian Brothers St., 375036 Yerevan, Armenia \label{Yerevan} \and
Department of Physics, Konan University, 8-9-1 Okamoto, Higashinada, Kobe, Hyogo 658-8501, Japan \label{Konan} \and
Kavli Institute for the Physics and Mathematics of the Universe (WPI), The University of Tokyo Institutes for Advanced Study (UTIAS), The University of Tokyo, 5-1-5 Kashiwa-no-Ha, Kashiwa, Chiba, 277-8583, Japan \label{KAVLI} \and
RIKEN, 2-1 Hirosawa, Wako, Saitama 351-0198, Japan \label{RIKKEN}
}